# Science goals and new mission concepts for future exploration of Titan's atmosphere, geology and habitability: Titan POlar Scout/orbitEr and *In situ* lake lander and DrONe explorer (POSEIDON)

Sébastien Rodriguez[1,*], Sandrine Vinatier[2], Daniel Cordier[3], Gabriel Tobie[4], Richard K. Achterberg[5], Carrie M. Anderson[6], Sarah V. Badman[7], Jason W. Barnes[8], Erika L. Barth[9], Bruno Bézard[2], Nathalie Carrasco[10], Benjamin Charnay[2], Roger N. Clark[11], Patrice Coll[12], Thomas Cornet[13], Athena Coustenis[2], Isabelle Couturier-Tamburelli[14], Michel Dobrijevic[15], F. Michael Flasar[6], Remco de Kok[16], Caroline Freissinet[10], Marina Galand[17], Thomas Gautier[10], Wolf D. Geppert[18], Caitlin A. Griffith[19], Murthy S. Gudipati[20], Lina Z. Hadid[21], Alexander G. Hayes[22], Amanda R. Hendrix[11], Ralf Jauman[23], Donald E. Jennings[6], Antoine Jolly[12], Klara Kalousova[24], Tommi T. Koskinen[19], Panayotis Lavvas[3], Sébastien Lebonnois[25], Jean-Pierre Lebreton[26], Alice Le Gall[10], Emmanuel Lellouch[2], Stéphane Le Mouélic[4], Rosaly M. C. Lopes[20], Juan M. Lora[27], Ralph D. Lorenz[28], Antoine Lucas[1], Shannon MacKenzie[28], Michael J. Malaska[20], Kathleen Mandt[28], Marco Mastrogiuseppe[29], Claire E. Newman[30], Conor A. Nixon[6], Jani Radebaugh[31], Scot C. Rafkin[9], Pascal Rannou[3], Ella M. Sciamma-O'Brien[32], Jason M. Soderblom[33], Anezina Solomonidou[34], Christophe Sotin[4], Katrin Stephan[35], Darrell Strobel[36], Cyril Szopa[10], Nicholas A. Teanby[37,*], Elizabeth P. Turtle[28], Véronique Vuitton[38], Robert A. West[20]

[*]Corresponding authors (rodriguez@ipgp.fr, N.Teanby@bristol.ac.uk)

[1]Université de Paris, Institut de physique du globe de Paris, CNRS, F-75005 Paris, France

[2]LESIA, Observatoire de Paris, Université PSL, CNRS, Sorbonne Université, Université de Paris, 5 place Jules Janssen, 92195 Meudon, France

[3]Groupe de Spectrométrie Moléculaire et Atmosphérique, UMR CNRS 7331, Université de Reims Champagne-Ardenne, Reims, France

[4]Laboratoire de Planétologie et Géodynamique, UMR 6112, CNRS, Université de Nantes, 2 rue de la Houssinière, Nantes 44322, France

[5]University of Maryland, Department of Astronomy, College Park, MD 20742, USA

[6]Planetary Systems Laboratory, NASA Goddard Space Flight Center, Greenbelt, MD 20771, USA

[7]Department of Physics and Astronomy, University of Leicester, Leicester, UK

[8]Department of Physics, University of Idaho, Moscow, Idaho, USA

[9]Southwest Research Institute, Boulder, Colorado, USA

[10]LATMOS/IPSL, UVSQ Université Paris-Saclay, Sorbonne Université, CNRS, Paris, France

[11]Planetary Science Institute Colorado, 1546 Cole Blvd #120, Lakewood, CO 80401, USA

[12]LISA, Université Paris-Est, Creteil, France.

[13]Aurora Technology BV for ESA - European Space Agency, European Space Astronomy Centre (ESAC), Camino Bajo del Castillo s/n, 28692 Villanueva de la Cañada, Madrid, Spain

[14]Aix-Marseille Université, CNRS, PIIM, UMR 7345, 13013 Marseille, France

[15]Laboratoire d'Astrophysique de Bordeaux, Université Bordeaux, CNRS, B18N, allée Geoffroy Saint-Hilaire, Pessac 33615, France

[16]Department of Physical Geography, Utrecht University, Utrecht, Netherlands

[17]Department of Physics, Imperial College London, Prince Consort Road, London SW7 2AZ, UK

[18]Department of Physics, AlbaNova University Center, Stockholm University, Roslagstullsbacken 21, Stockholm SE-10691, Sweden

[19]Lunar and Planetary Laboratory, University of Arizona, 1629 E. University Blvd., Tucson, AZ 85721, USA

[20]Jet Propulsion Laboratory, Caltech, 4800 Oak Grove Drive, Pasadena, CA 91109, USA

[21]Laboratoire de Physique des Plasmas (LPP), CNRS, Observatoire de Paris, Sorbonne Université, Université Paris Saclay, Ecole polytechnique, Institut Polytechnique de Paris, 91120 Palaiseau, France

[22]Cornell University, Ithaca NY, USA

[23]Institute of Geological Sciences, Free University Berlin, Germany

[24]Charles University, Faculty of Mathematics and Physics, Department of Geophysics, Prague, Czech Republic

[25]Laboratoire de Météorologie Dynamique (LMD/IPSL), Sorbonne Université, ENS, PSL Research University, Ecole Polytechnique, Institut Polytechnique de Paris, CNRS, Paris, France

[26]Laboratoire de Physique et Chimie de l'Environnement et de l'Espace (LPC2E), UMR 7328 CNRS, Université d'Orléans, France

[27]Department of Earth and Planetary Sciences, Yale University, 210 Whitney Avenue, New Haven, CT 06511, USA

[28]Johns Hopkins University Applied Physics Laboratory, 11100 Johns Hopkins Rd., Laurel, MD 20723, USA

[29]"La Sapienza" University of Rome, Italy

[30]Aeolis Research, 333 N. Dobson Road, Unit 5, Chandler, AZ 85224, USA

[31]Department of Geological Sciences, Brigham Young University, S-389 ESC Provo, UT 84602, United States

[32]NASA Ames Research Center, Space Science and Astrobiology Division, Moffett Field, CA 94035, USA

[33]Department of Earth, Atmospheric and Planetary Sciences, Massachusetts Institute of Technology, Cambridge, MA, USA

[34]Division of Geological and Planetary Sciences, California Institute of Technology, Pasadena, CA, USA

[35]DLR, Institute of Planetary Research, Berlin, Germany

[36]The Johns Hopkins University, Baltimore, MD, USA

[37]School of Earth Sciences, University of Bristol, Wills Memorial Building, Queens Road, Bristol, BS8 1RJ, UK

[38]Univ. Grenoble Alpes, CNRS, IPAG, 38000 Grenoble, France

**Abstract**

In response to ESA's "Voyage 2050" announcement of opportunity, we propose an ambitious L-class mission to explore one of the most exciting bodies in the Solar System, Saturn's largest moon Titan. Titan, a "world with two oceans", is an organic-rich body with interior-surface-atmosphere interactions that are comparable in complexity to the Earth. Titan is also one of the few places in the Solar System with habitability potential. Titan's remarkable nature was only partly revealed by the Cassini-Huygens mission and still holds mysteries requiring a complete exploration using a variety of vehicles and instruments. The proposed mission concept POSEIDON (Titan POlar Scout/orbitEr and *In situ* lake lander DrONe explorer) would perform joint orbital and *in situ* investigations of Titan. It is designed to build on and exceed the scope and scientific/technological accomplishments of Cassini-Huygens, exploring Titan in ways that were not previously possible, in particular through full close-up and *in situ* coverage over long periods of time. In the proposed mission architecture, POSEIDON consists of two major elements: a spacecraft with a large set of instruments that would orbit Titan, preferably in a low-eccentricity polar orbit, and a suite of *in situ* investigation components, i.e. a lake lander, a "heavy" drone (possibly amphibious) and/or a fleet of mini-drones, dedicated to the exploration of the polar regions. The ideal arrival time at Titan would be slightly before the next northern Spring equinox (2039), as equinoxes are the most active periods to monitor still largely unknown atmospheric and surface seasonal changes. The exploration of Titan's northern latitudes with an orbiter and *in situ* element(s) would be highly complementary in terms of timing (with possible mission timing overlap), locations, and science goals with the upcoming NASA New Frontiers Dragonfly mission that will provide *in situ* exploration of Titan's equatorial regions, in the mid-2030s.

## 1. Context and summary of a new mission's science goals and concepts

Saturn's largest moon, Titan, is one of the Solar System's most enigmatic bodies, as it is the only planetary moon that has a substantial atmosphere (with a column density larger than Earth's). Compared to the other planetary bodies, Titan is perhaps the one which most resembles Earth. Titan's atmosphere and surface are rich in organic material and set the stage for a complete meteorological cycle, involving surface liquids, rains, and storms, which is, however, based on methane ($CH_4$) as opposed to water on Earth. However, its methane-rich atmosphere is out of equilibrium as the ultraviolet radiation from the Sun irreversibly destroys methane, which should have disappeared within a few tens of millions of years. The $CH_4$ by-products, which end up forming a thick atmospheric haze, settle as organic sediments on Titan's surface. There are no obvious sources to supply the atmosphere with $CH_4$ except for the evaporation of the polar lakes. However, these lakes contain only a third of the total amount of $CH_4$ in Titan's atmosphere, and will be exhausted soon by geological time scales.

The entry of the Cassini-Huygens spacecraft into orbit around Saturn in July 2004 marked the start of a golden era in the exploration of Titan. While before this mission, huge hydrocarbon oceans dominated by $CH_4$ and $C_2H_6$ were expected to supply the atmospheric $CH_4$ (Sagan and Dermott 1982; Lunine et al. 1983), the Cassini prime mission (2004-2008) revealed the groundbreaking discoveries of the limited presence of liquids with dry equatorial dune fields (Lorenz et al. 2006) and lakes and seas mostly located at high northern latitudes, with a single lake near the south pole (Porco et al. 2005; Stofan et al. 2007). Another ground-breaking discovery was the detection of large positive and negative ions in the upper atmosphere above 1000km (Coates et al. 2007), with an increase in positive and negative ion and particle densities below 1000km (Waite et al. 2007; Shebanits et al. 2013, 2016), highlighting the complexity of chemistry forming aerosol precursors much higher than previously anticipated (Lavvas et al. 2013). The *in situ* Cassini instrument INMS (Ion and Neutral Mass Spectrometer) was not designed for such large and complex ions and their composition remains mostly unknown. In January 2005, the Huygens probe descended through Titan's atmosphere, taking the first close up pictures of the surface, revealing large networks of dendritic channels leading to a dried up seabed, and also obtaining detailed profiles of temperature and gas composition of the atmosphere at ≈10°S latitude (Tomasko et al. 2005; Fulchignoni et al. 2005; Niemann et al. 2005, 2010). The prime mission was extended with the Equinox mission (2008-2010) and then the Solstice mission (2010-2017), with a total of 127 targeted flybys of Titan. This brought new discoveries, mostly related to seasonal variations. In the atmosphere, a complete reversal of global atmospheric dynamics and the seasonal migration of clouds from southern to northern polar regions were observed as Titan moved from southern Summer to northern Summer solstice (Teanby et al. 2012; Vinatier et al., 2015; Rodriguez et al. 2011; Turtle et al. 2018; Coustenis et al. 2020). At the end of the Cassini mission, just after the Summer solstice, the global atmospheric dynamics had not stabilized to a state symmetrical to what was observed in the middle of the northern Winter, at the beginning of the mission (Coustenis et al. 2020; Coustenis 2021). Over the mission, rainfall events were recorded near the south pole in 2005 (during southern Summer) (Turtle et al., 2009), near the equator close to the northern Spring equinox in 2009 (Turtle et al. 2011c; Barnes et al. 2013) and in the northern polar region close to the Summer solstice (or northern Summer) (Dhingra et al., 2019), following the seasonal solar insolation. In the upper atmosphere, the Cassini mission revealed the high variability of the mesosphere (in terms of temperature and composition) with no correlation with solar radiation nor properties of Saturn's magnetic field. Wave dissipation may be a significant source of heating or cooling at those altitudes (Snowden and Yelle 2014). The extended missions (Equinox and Solstice) also allowed better characterization of Titan's surface nature and dynamics, and internal structure. A global geomorphological map has been drawn revealing the diversity of Titan's landscapes (extensive dunes, seas, filled and dried lakes, rivers, canyons, deltas, mountains, labyrinth terrains, and craters) (Lopes et al. 2020), indicative of the varying climatic (and erosive – aeolian, fluvial and/or chemical) environments that exist on Titan, from arid to more humid from equator to the poles. Only a few surface and near-surface changes have been identified yet: a few occurrences of surface darkenings due to seasonal rainfalls (Turtle et al. 2009, 2011c; Barnes et al. 2013; Dhingra et al. 2019), possible close-surface methane fog above or near the lakes in Summer in the polar regions (Brown et al. 2009), dust storms at the equator close to the Spring equinox, which are possibly associated with the activity of underlying dunes (Rodriguez et al. 2018; Karkoschka et al. 2019), waves, surface changes, and possible shoreline retreat at polar lakes as Summer was approaching (Turtle et al. 2011b; Barnes et al. 2014; Hofgartner et al. 2014; Cordier et al. 2017). Recent work has suggested that the observable surface is likely dominated by solid organic materials derived from atmospheric photochemistry that make their way to the surface and are further physically processed with occasional patches of icy bedrock (e.g. Rannou et al., 2015; Malaska et al. 2016, 2020; Solomonidou et al. 2018, 2020a; Brossier et al. 2018; Griffith et al. 2019), but this is still a subject of intense debate. The longevity of the Cassini mission also led to new insights into the structure and composition of Titan's interior and the exchange processes with the surface and atmosphere, confirming notably the presence of a global, subsurface salt-containing ocean (e.g. Mitri et al. 2014a).

Post-Cassini, we are now able to look back on the high-level scientific questions from the beginning of the mission, and assess the progress that has been made towards answering them. At the same time, new important scientific questions regarding Titan have emerged from the discoveries that have been made to date. Cassini was a dedicated mission to study the entire Saturn system and the limited number of Titan's flybys did not have a sufficient frequency (on average one flyby per month) to monitor atmospheric processes varying within a few hours/days (like variability in the thermosphere or clouds in the deep atmosphere, **Sections 2.1 and 2.3**) and to totally untangle the complexity of its surface dynamics and interior structure (aeolian and fluid transport, formation and evolution of lakes and seas, erosion, cryovolcanic activity, depth and thickness of its ice shell and global ocean, **Section 3**). Regarding the atmosphere, the region between 500km and 1000km, informally called the "agnostosphere", remains poorly known, as it was only probed *in situ* by Huygens (at a single location and time) and could only be studied using the few solar/stellar occultations measurements by the UV spectrometer (UVIS, Esposito et al. 2004). Cassini *in situ* measurements above 1000km altitude revealed a complex chemistry in the upper atmosphere (see **Section 2.1**), but only the smallest ionic species (< 100amu) could be tentatively identified by the Ion and Neutral Mass spectrometer (INMS, Waite et al. 2004), whereas the presence of large amounts of anions with higher masses up to 10,000amu/q were detected by the Cassini Plasma Spectrometer (CAPS, Young et al. 2004). Further, Cassini only indirectly constrained the global atmospheric circulation in the lowest altitude regions below 500km (**Sections 2.2 and 2.3**) (with the exception of the wind speed measurements of the Huygens probe during its descent near the equator). Many aspects of the complexity of the climatology of the moon in the lowest part of its atmosphere are also far from being fully understood, such as the distribution and seasonality of cloud formation, the vertical profiles of minor species in the deep atmosphere, the intensity of methane evaporation and precipitation, the origin and impact of atmospheric waves, or the intensity and direction of surface winds (**Section 2.3**). In the same manner, fundamental questions remain regarding Titan's surface and interior. To name a few, the age of Titan's surface is still poorly constrained (**Section 3.4**), the depth and thickness of its ice shell and global ocean are still unknown, and past or present cryovolcanic activity has still not been confirmed (**Section 3.5**). Its absolute surface reflectivity and composition are almost completely unknown (relevant to all topics of **Section 3**), the origin and morphodynamics of dunes, dissected plateaus, plains, rivers, lakes and seas, and associated erosion rates, are still strongly debated (**Sections 3.1, 3.2 and 3.3**). Finally, with respect to some definitions for habitability (presence of a stable substrate, available energy, organic chemistry, and the potential for holding a liquid solvent – Coustenis et al. 2013), Titan is one of the celestial bodies in the Solar System with the highest potential for habitability. Even if Cassini-Huygens provided essential observations to sustain this hypothesis, the real habitability potential of Titan is still strongly debated (which locations are best: deep surface, or surface? what are the source(s) of energy and where are they? how complex is Titan's organic chemistry? what are the solvents that are involved in Titan's chemistry? if water, how accessible is it from the surface? **Section 4**). We will not have more information on these fundamental questions without a new and dedicated planetary flight mission entirely devoted to exploring Titan, both at global and local scales.

In this article, we present a cross-sectional perspective of important scientific questions that remain partially or completely unanswered (see also Nixon et al. 2018), ranging from Titan's exosphere to its deep interior (considering Titan's atmosphere, surface, and interior as a complex interacting system, including the question of Titan's habitability), and we detail the necessary instrumentation and mission operational scenarios that can answer them. Our intention is to formulate the science goals for the next generation of planetary missions to Titan in order to prepare for the future exploration of the moon. We list here the primary questions that will be addressed by the proposed mission scenarios and payloads that optimize the intersection between all the needed instrumentation:

- **Science goal A: Titan's atmosphere**
    - **Upper atmosphere dynamics and chemistry:** What is the role of ion-neutral and heterogeneous chemistry in the upper atmosphere, including the chemistry involving organic aerosols? What drives its dynamics and what are the effects on ion densities? What are the physical processes in the ionosphere and thermosphere that drive ion densities and temperature variability? We identify here the need for an *in situ* high-resolution ion and neutral mass spectrometer, heterodyne submillimeter and UV spectrometers, a magnetometer, a pressure gauge, a Langmuir probe, a mutual impedance probe, and an electron spectrometer. Details are given in **Section 2.1**.
    - **Middle atmosphere dynamics and chemistry:** What generates Titan's atmospheric superrotation and what maintains it? How do the polar vortices form, evolve, and dissipate? What is the chemistry and its highest complexity attained inside polar vortices and what are the composition and structure of the massive stratospheric polar clouds? What are the composition, optical properties, and spatial distribution of aerosols in the main haze layer? We advocate here the need for an orbiter with a visible and near-IR imager, a heterodyne submillimeter, UV and far- to mid-IR spectrometers, a radio occultation experiment, orbital and *in situ* wind measurement experiments, an *in situ* high

resolution ion and neutral mass spectrometer, an imager/spectral radiometer, and a nephelometer/particle counter. Details are given in **Section 2.2**.

- **Lower atmosphere dynamics and methane cycle:** What are the characteristics of the $CH_4$ cycle on Titan? How do Titan's low atmospheric clouds form and evolve? What is the resulting precipitation rate? What is the wind regime near the surface? What is the composition of aerosols and how does it evolve through sedimentation and at the surface? Again, we promote here the need for an orbiter with a far- and mid-IR spectrometer and a submillimeter spectrometer, a radio occultation experiment, a visible near-IR imager, an orbital and *in situ* near-IR spectrometer, an *in situ* high resolution ion and neutral mass spectrometer, a nephelometer/particle counter, and a camera/spectral radiometer. Details are given in **Section 2.3**.

- **Science goal B: Titan's geology**
  - **Aeolian features:** What is the precise morphometry of Titan's dunes? Do they change over observable timescales? How are they located with respect to other landscapes? What is the total volume of solid organic sediment trapped in the bedforms? What are the nature (composition, grain size of the sand material), origin (source of sand), and dynamics (connection with modern winds, mode and rate of growth) of the dunes? This can be answered by the use of orbited and *in situ* high-resolution multi-wavelength remote sensing packages (including cameras, spectrometers, and spectral-imagers) and *in situ* surface material sampling and analysis capabilities. Details are given in **Section 3.1**.
  - **Incision and chemical erosion:** What are the location, morphology, and spatial scale of Titan's river networks? What are the associated formation processes and erosion rates? Are all rivers still active? Do all river properties and formation mechanisms vary with latitude and local climatic conditions? What is the thickness of the labyrinth terrain deposits? Is there evidence of layering in these deposits? At what thickness? Do the compositions of the depositional layers change? This can be thoroughly studied with the help of a long-lived Titan orbiter and a high-resolution remote sensing package (imagery and spectral-imagery down to decameter scale) and a mobile *in situ* probe with remote sensing instruments and sampling capabilities. Details are given in **Section 3.2**.

- **Seas and lakes:** What are the distribution, shapes, and the precise composition of Titan's seas and lakes, down to the sub-kilometer scale? How do they connect with the hydrological network? How much liquid hydrocarbon is stored in the seas and lakes? How do seas and lakes form and change with seasonal and inter-seasonal timescales? A long-lived Titan orbiter with a near-polar orbit will be required, including a Synthetic Aperture Radar (SAR) system, a Ground Penetrating Radar system, and/or a high-precision altimeter. An *in situ* mobile/floating/submarine probe, including a spectral-imager, electrical environment and meteorological packages, and sampling capabilities would provide a fundamental support to those questions. Details are given in **Section 3.3**.
- **Craters and Mountains:** What is Titan's crater distribution, down to the decameter scale? What is the distribution and orientation of mountain blocks and ridges across Titan, and how do they correlate with other global topography? What are the relative ages of Titan's geomorphological units? What is the bedrock/crust composition? What is the erosion rate of craters and how does it change with location on Titan? A long-lived orbiter, with high-resolution imaging capabilities (down to 10 meters), would allow a global survey of impact and mountain/tectonic features. Details are given in **Section 3.4**.
- **Internal structure and subsurface ocean:** What are the depth, volume, and composition of the subsurface liquid water ocean? Is Titan currently, or has it been in the past, cryovolcanically active? Are there chemical interactions between the ocean, the rock core and the organic-rich crust? How did Titan's atmosphere form and evolve with time in connection with the interior? A combination of geophysical measurements from the orbit (radio experiment, radar altimeter and sounder, radar imager, magnetometer and plasma package) and from the ground (seismometer, radio transponder, electric sensors, and magnetometer) is required to constrain the interior and hydrosphere structures. Measurement by a high-precision mass spectrometer of the isotopic ratio in atmospheric noble gases will also put fundamental constraints on how water-rock interaction have occurred in Titan's interior. A future mission with a high-resolution microwave radiometer on board could search for thermal anomalies (hot spots) possibly revealing current active cryovolcanism on Titan. Details are given in **Section 3.5**.

- **Science goal C: Titan's habitability**
    - How is the organic material falling from the atmosphere physically/chemically processed at the surface? What is the nature of dissolved species in hydrocarbon lakes? Does this liquid environment harbor a chemical reaction network? What is the nature and quantity of material exchange between the subsurface ocean and the surface? In the past, did a form of life develop in water ponds formed by cryovolcanism or bolide impacts? Is there evidence for deep ocean materials, including potential biosignatures, having been extruded and deposited onto the surface? A very high-resolution mass spectrometer should be used for low atmosphere composition measurements. An instrument of the same class is also needed for the *in situ* analysis of liquid phases and solid surfaces, and subsurface sample analysis complemented by specific samplers for both phases, possibly associated with a drill (down to 50 cm to 1 m), searching for water ice. Details are given in **Section 4**.

Along with standard, but also with new generation instrumentation, key new instruments, which were not aboard the Cassini mission, are essential to answer a wealth of the preceding questions, especially a very high-resolution mass spectrometer, a ground penetrating radar on an orbiter and a probe/drone, and an orbiter submillimeter heterodyne spectrometer. The required measurements to answer the open questions cannot be performed from terrestrial ground-based or space-borne facilities.

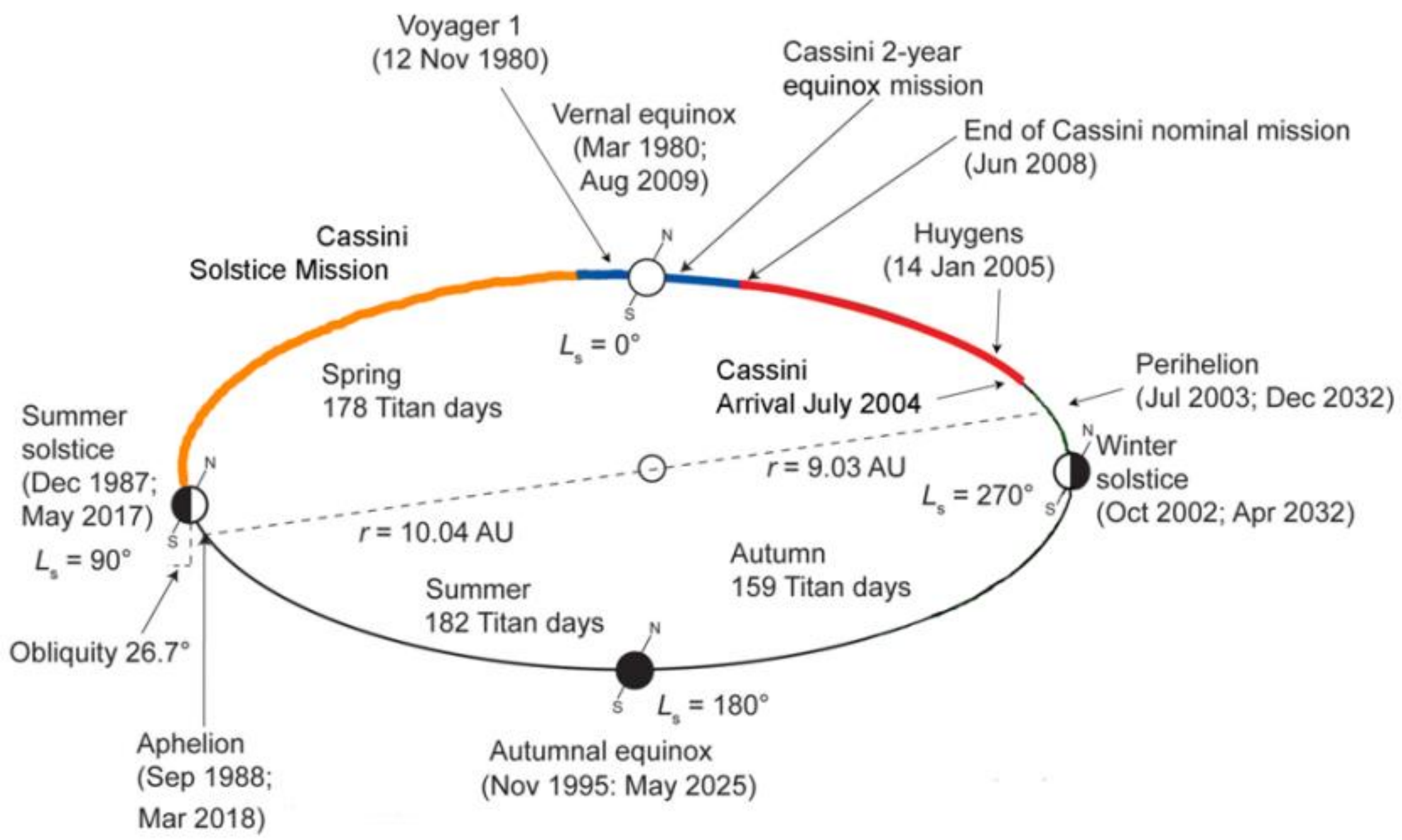

*Figure 1: Titan experiences long seasons of about 7.5 Earth years. A mission arriving in 2030-2040s will encounter similar seasons to Cassini-Huygens allowing for data set inter-comparison and evaluation of interannual changes. This will also provide an excellent complement to the Dragonfly mission that should arrive at Titan by 2034 (Lorenz et al. 2017). (From "Titan Explorer: Flagship Mission Study" by J. C. Leary et al., Jan 2008.)*

The ESA L-class mission concept that we propose here consists of a Titan orbiter and at least one *in situ* element (lake lander and/or drone(s)). The choice of an orbiter would guarantee global coverage, good opportunities for repeat observations to monitor atmospheric and surface changes, and high spatial resolution for Titan's atmosphere and surface observations both in imagery and spectroscopy. Ideally, the orbiter would have a high inclination elliptical orbit whose closest approach altitude will be localized in the thermosphere to perform *in situ* mass spectrometry measurements at each orbit. A – preferably mobile – *in situ* package (drone(s)) would be dedicated to study Titan's areas of particular geological interest at unprecedented coverage and spatial resolution. It would also be used to perform atmospheric measurements during the probe descent, in the lowest portion of the atmosphere during flights (for an aerial probe), and at the surface. The *in situ* element(s) should preferentially be sent to high northern latitudes to study the lake/sea region and to perform atmospheric measurements inside the polar vortex. The ideal arrival time at Titan would be slightly before 2039 (the next northern Spring equinox, while the northern Autumn equinox will not occur before 2054), as equinoxes are the most exciting observing periods to monitor the most striking and still largely unknown atmospheric and surface seasonal changes (impacting e.g. stratospheric chemistry and dynamics, geographic distribution and intensity of cloud activity and rainfall, near-surface wind strength and direction; Teanby et al. 2012; de Kok et al. 2014; Vinatier et al. 2015, 2020; Rodriguez et al. 2009, 2011, 2018; Turtle et al. 2018; Charnay et al. 2015; Coustenis 2021). In addition, a mission arriving in the 2030s will encounter a season similar to that during the Cassini-Huygens mission allowing for dataset inter-comparison and evaluation of interannual changes (**Figure 1**).

With an arrival slightly before 2039, the presence of an orbiter and the exploration of Titan's northern latitudes would be highly complementary in terms of timing (with possible mission timing overlap), locations, and science goals with the upcoming NASA New Frontiers Dragonfly mission that will study equatorial regions *in situ* with a planned arrival by 2034. Indeed, the New Frontier scope and architectural choices that make Dragonfly best suited for its local *in situ* investigation necessarily preclude addressing many other outstanding questions at Titan, especially those requiring a global perspective[1] (MacKenzie et al. 2021) . In particular, exploring *in situ* the polar lakes and seas, their influence on Titan's global hydrologic cycle, and their potential habitability, will remain out of Dragonfly's range. Such measurements could also be complemented by orbital imaging at higher spatial and temporal resolutions than what Cassini or ground-based observations could provide. A higher-order gravity field might reveal eroded craters and thus constrain the prevalence of transient liquid water environments. More specifically, Dragonfly's seismic investigation of the interior would be significantly enhanced by a global topographic dataset and higher fidelity mapping of the gravity field. Further study of Titan's climate and the seasonal evolution of hazes and weather phenomena

[1] MacKenzie et al., "Titan: Earth-like on the Outside, Ocean World on the Inside", NASEM Decadal Survey 2023-3032.

(e.g. clouds and haboobs, Rodriguez et al. 2011, 2018; Turtle et al. 2018) requires continued long-term monitoring from orbit. Global imaging and spectral-imaging datasets would facilitate understanding the beginning-to-end life cycle and transport of the surface materials sampled and observed *in situ* by Dragonfly. Even better science return from Dragonfly and a companion orbiter would be obtained if the timing overlap could occur around 2039.

Note that a mission including an orbiter (and possibly an *in situ* element such as the Titan Saturn System Mission – Reh et al. 2008) arriving outside the Dragonfly mission calendar, or equinoxes, would still have an outstanding scientific impact, complementing the drone results and answering fundamental open questions that remain about Titan's system that cannot be answered from terrestrial ground-based or space-borne facilities.

## 2. Science Goal A: Titan's atmosphere

This section will first cover questions related to the complex chemistry, the high variability, and the dynamics of the upper atmosphere. We will then present the main questions related to middle atmosphere dynamics, including superrotation and polar vortices, and chemistry in this deeper region including the haze composition. We will then focus on the lower part of the atmosphere with questions related to the $CH_4$ cycle, the cloud formation and seasonal evolution, and the wind regime close to the surface.

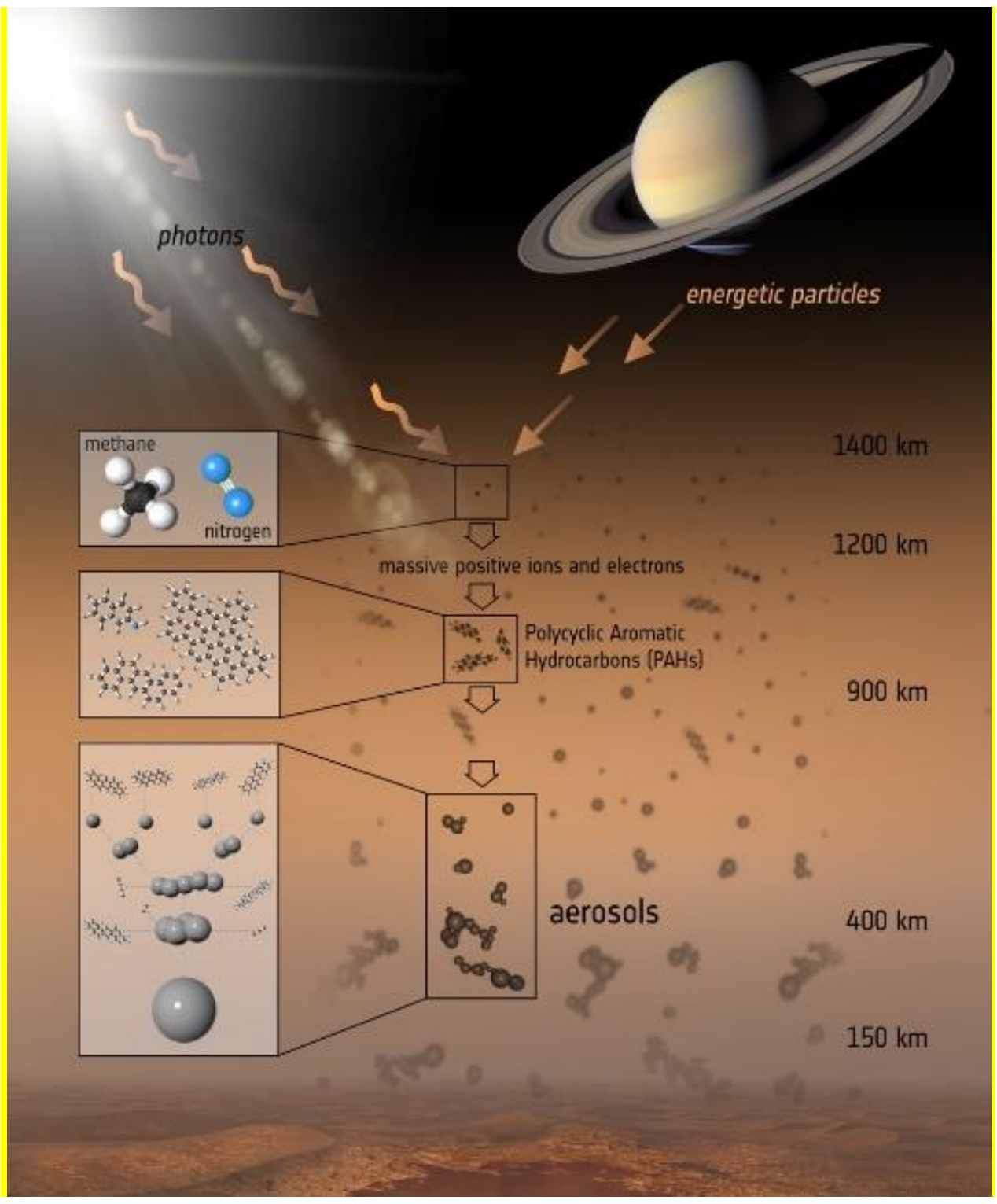

*Figure 2: Titan's atmosphere consists of the two primary gases, methane ($CH_4$) and nitrogen ($N_2$), which undergo a series of photochemical reactions to produce heavier molecules, and the ubiquitous haze particles (aerosols). (ESA/ATG medialab)*

Titan's atmosphere contains one of the most complex chemistries in the Solar System. Micrometer-size aerosols are formed from a chain of chemical reactions initiated in the upper atmosphere by ionization and dissociation of nitrogen and methane by solar UV photons and associated photoelectrons (Galand et al. 2010), see **Figure 2**. The photochemically produced molecules and aerosols have a strong impact on the radiative budget of Titan's atmosphere, and consequently on its climate. Transport by global dynamics, which reverses each half Titan-year, greatly affects the distribution of these compounds. The complex couplings of chemistry, radiation, and dynamics make Titan's atmosphere an ideal laboratory to understand physical and chemical processes at play in atmospheres and particularly those showing the presence of photochemical haze such as Pluto (Cheng et al. 2017), or on any of the increasing number of exoplanets (Pinhas et al. 2019). Furthermore, the study of Titan's chemistry has strong astrobiological implications as it naturally produces complex nitrogen-containing organic molecules, which can act as biomolecule precursors, including formation of the DNA base adenine. Titan's atmosphere also contains oxygen compounds, which could play a role in the formation of amino-acids and oxygen-containing DNA nucleobases, as suggested by laboratory experiments that simulate Titan's chemistry (Hörst et al. 2012).

## 2.1 The upper atmosphere: a region of complex physical and chemical processes

### *2.1.1 Current knowledge*

One of the most outstanding observations of the Cassini mission was the *in situ* detection of ions with several hundred mass units above 900km (Coates et al. 2007; Waite et al. 2007; Shebanits et al. 2013, 2016) revealing the dusty plasma nature of Titan's upper atmosphere (located above 550km altitude) and its highly complex ion-neutral chemistry (Vuitton et al. 2006, 2007) producing nanoparticles at much higher altitude than previously thought (Lavvas et al. 2013). On Titan's dayside, the main source of ionization is the solar extreme ultraviolet (EUV) radiation (Ågren et al. 2009; Edberg et al. 2013) while ionization from energetic electrons from Saturn's magnetosphere contributes as a source of night side short-lived ions (Cui et al. 2009; Sagnières et al. 2015; Vigren et al. 2015). In addition, day-to-night transport from the neutral atmosphere seems to be a significant source of the night side long-lived ions (Cui et al. 2009). Titan's thermosphere shows an unexpectedly variable $N_2$ (Titan's main gaseous species) density that changes by more than an order of magnitude on relatively short timescales (comparable to a few or less Titan days), accompanied by large wave-like perturbations in vertical thermal structure (Westlake et al. 2011; Snowden et al. 2013). Titan, similarly to Venus, possesses an atmosphere in global superrotation, which extends up to at least 1000km with a surprisingly high wind speed of 350m/s (Lellouch et al. 2019). At altitudes of 500-1000km, aerosol seed particles grow through coagulation and chemistry (Lavvas et al. 2011a) while drifting downward into the deep atmosphere. Increasingly complex molecules are also produced by neutral photochemistry (Loison et al., 2019; Vuitton et al. 2019) and Cassini measurements, albeit limited in scope, revealed the density profiles of several key hydrocarbons and nitriles, the extinction due to aerosols and the possible presence of polycyclic aromatic hydrocarbons (PAHs) (Liang et al. 2007; Delitsky and McKay, 2010; Koskinen et al. 2011; Lopez-Puertas et al. 2013; Maltagliati et al. 2015; Dinelli et al., 2019; Cours et al., 2020).

### *2.1.2 Open questions*

Cassini only partially revealed the role of ion-neutral chemistry in Titan's upper atmosphere and could not address the chemical nature of heavy neutrals and ions and their formation mechanisms, or the identity of macromolecules because of the limited mass range and resolution of Cassini's *in situ* mass spectrometers (INMS had a mass resolution of m/Δm<500 at *m/z* 50). However, ion-neutral chemistry may dominate the formation of larger molecules above 900km (Lavvas et al. 2011b). These nanoparticles constitute only 10% of the total haze mass flux observed in the lower atmosphere. Thus, further chemical growth based on neutral chemistry needs to take place below the ionosphere through heterogeneous processes on the nanoparticle surface, but the details of this mechanism are unknown. More generally, Titan's upper atmosphere offers the perfect natural laboratory to assess complex ion-neutral chemistry and to characterize dusty plasmas and heterogeneous chemical processes that give birth to photochemical hazes. Such hazes are encountered in other astrophysics environments, such as Pluto's atmosphere (Gao et al., 2017) and the plumes of Enceladus (Morooka et al. 2011; Hill et al. 2012), and are expected to be present in exoplanetary atmospheres (Helling, 2019). Thus, a better understanding of their formation will have important ramifications for multiple environments.

Titan's atmosphere contains oxygen compounds (CO, $H_2O$, $CO_2$) and $O^+$ ions possibly sourced from Enceladus (Hörst et al. 2008; Dobrijevic et al. 2014). Whether oxygen is incorporated into the more complex organic molecules on Titan remains an outstanding question with great implications for prebiotic chemistry. If oxygen is detected in organics in the upper atmosphere, it can imply the presence of a global flux of a richer diversity of prebiotic molecules descending to the surface.

The recent detection of strong thermospheric equatorial zonal winds with speeds increasing with height up to 350m/s at 1000km (**Figure 3**, Lellouch et al. 2019) was totally unexpected. The source of such rapid winds could be driven by instabilities on the flanks of the strong stratospheric winter hemisphere zonal jet or be related to waves produced in the stratosphere in response to the diurnal variation of the solar insolation and propagating toward the upper atmosphere. These gravity waves, which were observed by Cassini-Huygens (Fulchignoni et al. 2005) and in stellar occultations observed from the ground (Sicardy et al. 2006), could transfer momentum from the deepest atmospheric layers towards the upper atmosphere and accelerate equatorial winds. Monitoring the winds with latitude, local time, and season will give a clue to its origin. This will also be critical in order to assess the effect of high winds on the ion densities, especially on the night side where transport from dayside is essential for long-lived ions (Cui et al. 2009).

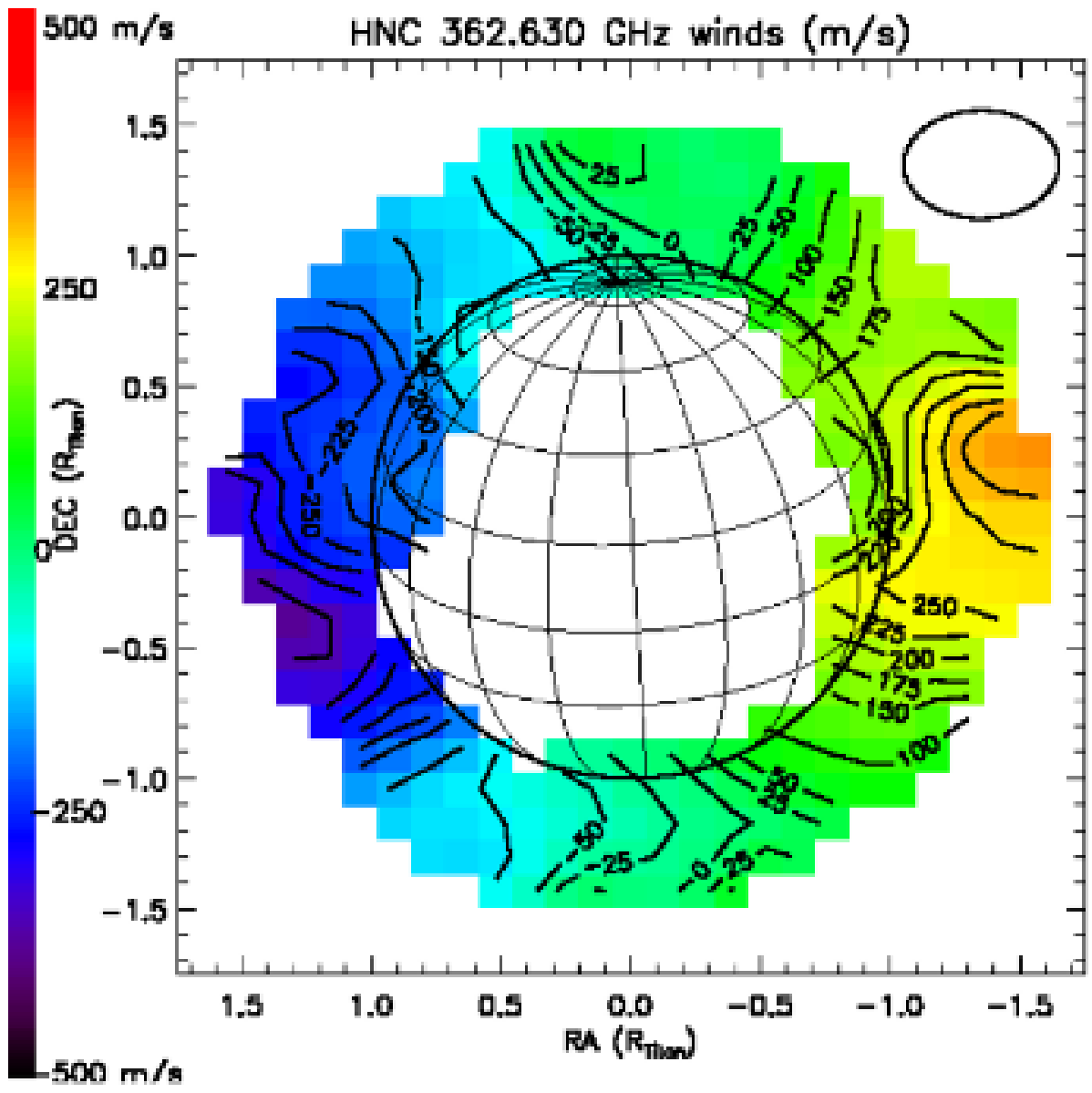

Figure 3: Wind speed maps measured from HNC line Doppler shift. Distance is expressed in Titan's radius. Blue color shows approaching winds and yellow-orange shows recessing winds. Winds measured from HNC are more localized around the equator with speeds as high as 350m/s (from Lellouch et al. 2019).

A global picture of how the upper atmosphere works could not be derived from the Cassini measurements alone because the flybys were too scarce to monitor the density and temperature variability that seems to occur on timescales smaller than one Titan day. No correlations were found between temperature, latitudes, longitudes, or solar insulation, which suggests that the temperature in this region is not controlled by the UV solar flux absorption. Understanding the origin of this variability is critically important because it affects the escape rates from Titan's atmosphere (Cui et al. 2011, 2012) and is likely to have consequences for the circulation and ion-neutral chemistry in the upper atmosphere.

To summarize, the main important questions related to the upper atmosphere are:

- **[1] What is the role of ion-neutral and heterogeneous chemistry in the upper atmosphere? What is the degree of oxygen incorporation in photochemical species?**
- **[2] What drives the dynamics of the upper atmosphere and what is its origin? What is the influence of the lower atmosphere and atmospheric waves generated there on the thermospheric circulation? What is the effect on ion densities?**
- **[3] What are the physical processes in the ionosphere and thermosphere that drive their density and temperature variability?**

*2.1.3 Proposed instrumentation and mission concept to address the open questions*

A critical scientific advancement compared to the Cassini mission with the proposed instrumental package will be *in situ* atmospheric sampling and remote sensing at altitudes much lower than those probed with Cassini.

**Understanding the upper atmospheric chemistry (open questions 1 & 2)** requires the identification of common subunits and building blocks that form the nanoparticles so that we can predict the chemical structures and reactivity of the large molecules. **This can only be achieved by using *in situ*** **analysis**. A new generation of ion and neutral mass spectrometer with a much higher mass resolution and a higher mass upper limit than INMS and CAPS aboard Cassini is absolutely required on a **Titan orbiter**. Such an instrument should be able to determine the composition of macromolecules up to 1000u with a mass resolution m/Δm = 100,000 at *m/z* 100 (**Figure 12**) and a sensitivity of $10^{-3}$ molecule.$cm^{-3}$. The CosmOrbitrap instrument (Briois et al. 2016), which is under development, is designed to meet these requirements.

**Understanding the variability of the thermosphere and the origin of supersonic winds (open questions 2 & 3)** in the upper atmosphere and how they evolve with seasons requires a **Titan orbiter** carrying:

- **A heterodyne submillimeter (sub-mm) spectrometer,** in order to directly measure the wind from the Doppler shifts of the numerous spectral lines from the surface up to at least 1200km with a vertical resolution of ≈10km (Lellouch et al. 2010). This will provide the first 3-D wind measurement on a celestial body other than the Earth. Such an instrument is also of prime interest for retrieving the thermal profile (from CO and HCN lines) and the mixing ratio profiles of many molecules ($H_2O$, $NH_3$, $CH_3C_2H$, $CH_2NH$, $HC_3N$, $HC_5N$, $CH_3CN$, $C_2H_3CN$, $C_2H_5CN$…) and their isotopes in the same altitude range, giving crucial new insights into the chemistry of the upper atmosphere (Loison et al., 2017; Dobrijevic et al., 2018). These observations would be a major advance compared to what was done by Cassini, which mostly probed the 100-500km region (from mid- and far-IR CIRS observations) and the ≈1000km region (*in situ* with INMS and CAPS), while only a few UV solar/stellar occultations probed the 500-1000km region. Here, it should be noted that the Atacama Large Millimeter Array (ALMA) will never provide a spatial resolution better than ≈400km on Titan's disc.
- **An ultraviolet (UV) spectrometer,** necessary to reveal the atmospheric density variability (from $N_2$), the trace gas density variability ($CH_4$, $C_2H_2$, $C_2H_4$, $C_4H_2$, $C_6H_6$, HCN, and $HC_3N$) and the aerosol extinction in the 500-1400km range from stellar and solar occultations observed by a **Titan orbiter**. It will also measure the temperature structure at 500-1400km from $N_2$ density determined from EUV range in the 1100-1400km region and indirectly at deeper altitude (above 600km) from the $CH_4$ density assuming that it is uniformly mixed with $N_2$. Observations of UV airglow and reflected light will constrain energy deposition and aerosol properties that will further contribute to evaluating the impact of hazes in the upper atmosphere energy balance and the growth rate of particles due to heterogeneous chemical processes. To overcome the limitations of Cassini observations, the measurements should be designed to achieve high vertical resolution and be targeted to map spatial and temporal trends in dedicated campaigns. Compared to Cassini/UVIS, the wavelength coverage should be extended to include the middle UV (MUV) range in addition to the EUV and FUV ranges. This would allow for better spectral constraints on aerosol extinction and possible detection of complex molecules below 1000km that were not accessible to Cassini.

**Understanding physical processes in the ionosphere (open questions 1, 2 & 3)** requires:

- **A pressure gauge,** providing absolute measurement of the total neutral density; such measurements would be combined with the neutral composition measurements from a high-resolution mass spectrometer (e.g. CosmOrbitrap, see above) to provide reliable absolute neutral densities. Such an instrument would also provide critical positive ion composition measurements to identify the source of long-lived ions and the role played by transport through the thermospheric winds.
- **A Mutual Impedance Probe (combined with a Langmuir Probe),** which provides absolute electron densities to be compared with the total positive ion densities from the positive ion mass spectrometer to highlight any difference between both in the region of dusty plasma. It would also provide electron temperatures relevant for assessing the energy budget and constraining ion-electron reaction coefficients for assessing ionospheric density.
- **A Negative Ion Mass Spectrometer,** which measures the composition and densities of negative ions (which had not been anticipated ahead of the Cassini mission, resulting in CAPS/ELS not being calibrated for such ions and not having suitable mass resolution) extending towards very high masses (as CAPS/ELS detected negative ions beyond 10,000u/q (Coates et al. 2007)). This will provide critically new insights on the negative ion chemistry and formation of multiple negative charged ions (Vuitton et al. 2009).

In addition to these new sensors not yet flown at Titan, we would need complementary observations to provide context and critical observations for multi-instrument studies in order to address the key questions:

- **A Langmuir probe,** measuring variability in electron density and temperature with high time resolution. It would also measure the spacecraft potential critical for interpreting the ion and electron spectrometer dataset.
- **An Electron Spectrometer** (1 eV-1 keV), which measures the electron densities as a function of energy. This is essential for assessing the energetic electron population ionizing the upper atmosphere and assessing the relative importance of local ionization versus transport from dayside for the source of ions on the night side. It would also be essential for constraining the photoelectron source on the dayside.
- **A Magnetometer,** which measures the magnetic field vector components and, combined with 3-D magnetospheric model, to derive the 3-D configuration of the magnetic field lines, along which particles are transported. This is essential information for assessing the electron energy budget (Galand et al. 2006) and the interaction of Titan with Saturn's magnetosphere (Snowden et al. 2013). Magnetic field measurements would also be used for deriving pitch angle information for the electrons measured by the electron spectrometer.

**Understanding the upper atmospheric chemistry (open questions 1, 2 & 3)** requires the identification of common subunits and building blocks composing the nanoparticles so we predict the chemical structures and reactivity of the large molecules. **This can only be achieved by using *in situ* analysis**. A high-resolution ion and neutral mass spectrometer (e.g. CosmOrbitrap) with a much higher mass resolution and a higher mass upper limit than INMS and CAPS aboard Cassini is absolutely required on a Titan orbiter.

It is important to emphasize that all the above required measurements can only be performed with a dedicated **Titan orbiter** as the Autumn/Winter polar region is unobservable from the Earth and any of the largest current or future ground-based (ALMA, ELT) or space-borne facilities (JWST) will not have enough spatial resolution to address the questions mentioned above.

## 2.2 The middle atmosphere: global dynamics and its coupling to composition and haze distributions

### *2.2.1 Current knowledge*

Titan's middle atmosphere is typically located above 80km, where most molecules condense (with the exception of CO, $H_2$, and $C_2H_4$) at Titan's average temperature and pressure conditions. This region is a transition between the upper atmosphere (**Section 2.1**), where molecules and aerosols are formed, and the troposphere (**Section 2.3**), in which convection controls surface-atmosphere interactions. Photochemical species' sources in the upper atmosphere and condensation sink in the deeper atmosphere result in increasing-with-height concentrations profiles, with vertical gradients strongly impacted by the global dynamics. The middle atmosphere contains the main haze layer with a permanent but altitude-variable detached haze layer on top during Summer/Winter. A significant correlation is observed between the detached haze layer position and a sharp transition in the temperature profile, marking the end of the mesosphere (and the boundary with the thermosphere) and the presence of large-scale gravity waves above (e.g. Porco et al. 2005; Fulchignoni et al. 2005).

Like on Earth, a polar vortex forms on Titan during Winter. Stratospheric polar vortices are regions of particular interest in planetary atmospheres. They are dominant dynamical structures, in which the air is isolated from the rest of the atmosphere by high-speed zonal winds and whose morphology strongly varies with season. On Titan, a strong enrichment of photochemically produced species is observed inside polar vortices (e.g., Coustenis et al. 2018; Mathé et al., 2020; Teanby et al. 2017, 2019; Vinatier et al. 2015, 2020), and massive polar stratospheric clouds with complex compositions have been detected during the northern Winter and Spring and during southern Autumn (Le Mouélic et al. 2012, 2018; de Kok et al. 2014; West et al. 2016; Anderson et al. 2018, Vinatier et al. 2018).

Atmospheric superrotation is intimately linked to the meridional circulation that shows marked seasonal changes with global pole-to-pole circulation during Winter/Summer and equator-to-pole circulation close to the equinoxes (every ≈15 years). This meridional circulation transports photochemical species (haze and molecules) that impact the radiative budget of the atmosphere and in turn affect the global dynamics. Aerosols especially strongly impact the radiative balance as they control the stratospheric temperature by diabatic heating in the visible and by dominating the cooling to space in the infrared especially during the Winter polar night (Rannou et al., 2004; Larson et al., 2015; Bézard et al. 2018). Large-scale aerosol structures result from interaction with the atmospheric circulation, such as the global thin detached haze layer whose altitude drastically changed with season (West et al. 2011, 2018). A sharp minimum of zonal wind around 70-80km altitude was observed in the lower stratosphere from Huygens *in situ* measurements (Bird et al. 2005) and indirectly derived from radio-occultation measurements (Flasar et al. 2013). This almost zero-wind layer seems to decouple the global dynamics in the deep and in the middle atmosphere.

Chemistry is also active in the middle atmosphere, while being less productive than in the upper atmosphere. Cassini/Composite InfraRed Spectrometer (CIRS, Flasar et al. 2004) and Visual and Infrared Mapping Spectrometer (VIMS, Brown et al. 2004) measurements provided a seasonal monitoring of the vertical and spatial distributions of a dozen species (e.g. Coustenis et al. 2013, 2020; Sylvestre et al. 2018; Teanby et al. 2019; Vinatier et al. 2010, 2015, 2020), constraining both 1-D photochemical models (e.g. Vuitton et al. 2019) and General Circulation Models (GCMs) (e.g. Lebonnois et al. 2012). The inventory of complex molecules was recently extended with the detection of $C_3H_6$ with CIRS (Nixon et al. 2013) and subsequently $C_2H_5CN$ and $C_2H_3CN$ with ALMA (Cordiner et al. 2015, 2019; Palmer et al. 2017), albeit with limited horizontal and vertical resolution. Those two latter molecules could not be observed by Cassini due to the limited spectral coverage and/or sensitivity of the instruments. Laboratory studies indicate continuous photopolymerization of unsaturated organic molecules in ice and aerosol phase resulting in aerosol growth as well as chemical aging caused by UV photons available at various altitudes (Carrasco et al. 2018; Couturier-Tamburelli et al. 2014, 2015, 2018; Gudipati et al. 2013). Such solid-phase photochemistry could continue on Titan's surface.

*2.2.2 Open questions*

GCMs currently favor scenarios involving planetary-scale barotropic wave activity in the Winter hemisphere (Newman et al., 2011; Read and Lebonnois 2018) to generate superrotation but some models have difficulties maintaining it, especially at equatorial latitudes. Possible signatures of these waves were detected recently (Cordier et al. private communication) on the haze spatial distribution from Cassini/Imaging Science Subsystem (ISS, Porco et al. 2004) images, but because of the limited number of flybys, only a few snapshots of their spatial and temporal evolutions, insufficient to constrain the models, are available. It is necessary to know how these waves evolve spatially on timescales from days to seasons to understand their impact on angular momentum transport and their role on generating and maintaining superrotation. This will be crucial to elucidate mechanisms at play in Titan's atmosphere and more generally for other partially superrotating atmospheres in the Solar System or for tidally-locked exoplanets (e.g. Pierrehumbert 2011). The minimum zonal wind near 80km altitude is currently not well-reproduced by GCMs and its consequences on angular momentum exchanges and transport of haze and trace species between troposphere and stratosphere are unknown.

During the Cassini mission, only a partial view of the vortex formation and its seasonal evolution was obtained, using (i) temperature and trace gas concentration distributions derived from CIRS (e.g. Achterberg et al 2011; Coustenis et al. 2018; Sharkey et al. 2020; Teanby et al. 2017, 2019;; Vinatier et al. 2015, 2020), (ii) seasonal evolution of massive stratospheric polar clouds first observed at the north pole during northern Winter and later at the south pole during southern Autumn (Jennings et al.

2012a,b; Le Mouélic et al. 2012, 2018; West et al. 2016; see **Figure 4**). Polar vortex structures change on quite rapid timescales: for instance, the southern polar vortex has doubled in size within a few Titan days in early Autumn. We currently do not know what controls the latitudinal extent of the polar vortex and how it forms and disappears. Its vertical structure, while it was forming at the south pole, cannot be inferred between 2012 and 2015 because of too few Cassini limb observations of polar regions. Polar vortices are regions of strong interaction with the upper atmosphere as the subsiding air comes from above but the vortex structure across the upper atmosphere and stratosphere cannot be extracted from the Cassini observations as no UV occultations occurred in this region during the vortex formation phase.

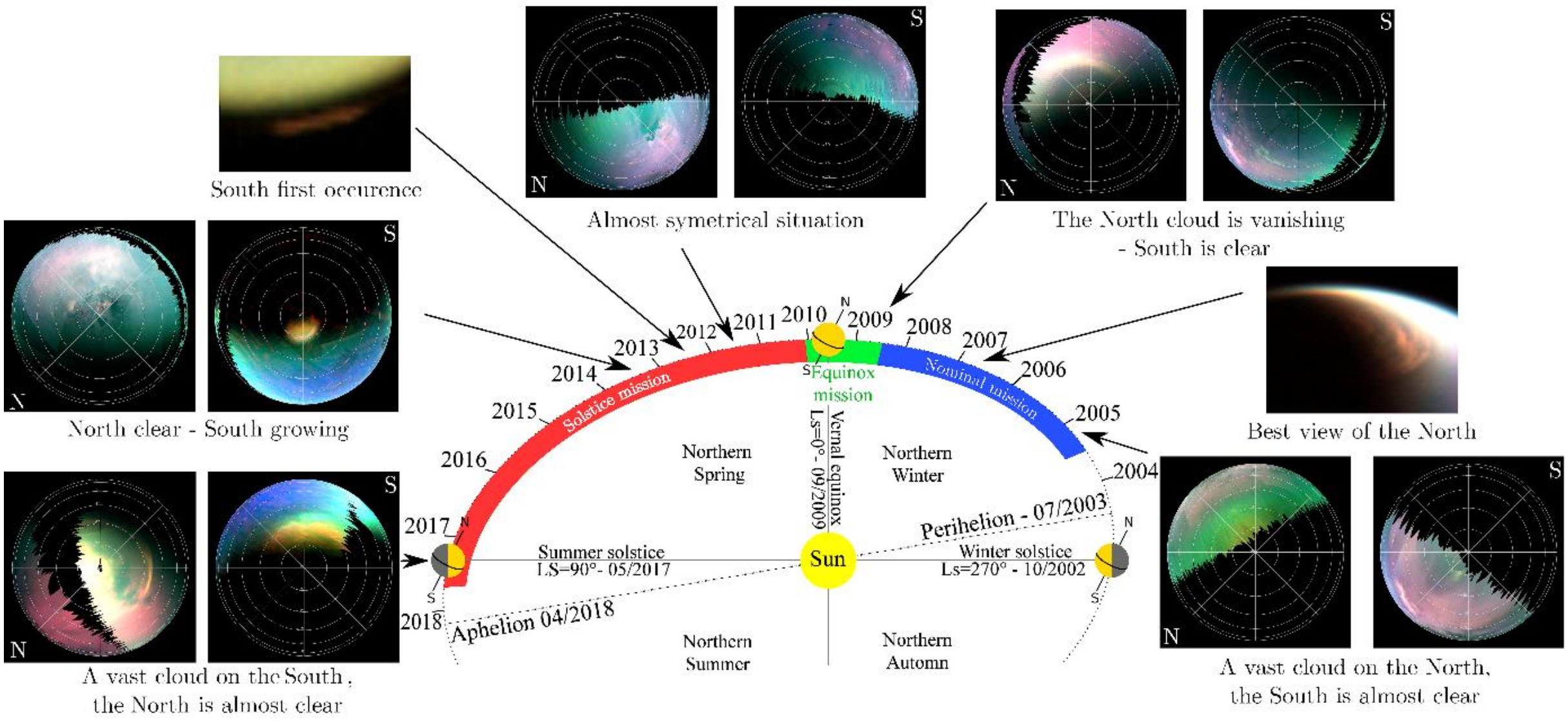

*Figure 4: Polar stratospheric cloud seasonal variations (from Le Mouélic et al. 2018).*

The chemistry inside Titan's vortices is particularly interesting, showing the strongest enrichments in photochemically produced species and thus potentially producing even more complex chemistry than elsewhere in the middle atmosphere. The neutral atmosphere exhibits rapid and dramatic changes with seasons. The highest degree in complexity of the organic chemistry is unknown. The high molecular concentrations combined with the low temperatures result in condensation of almost all molecules in the stratosphere as high as 300km altitude as observed at the south pole during Autumn and result in the production of the most complex stratospheric ice clouds observed in the Solar System. The composition of these massive clouds is currently poorly known. Only a few condensates have been identified: $HC_3N$ ice (Anderson et al. 2010), $C_4N_2$ ice (Anderson et al. 2016), HCN ice (de Kok et al. 2014; Le Mouélic et al. 2018), $C_6H_6$ ice (Vinatier et al. 2018), and co-condensed HCN:$C_6H_6$ ices (Anderson et al. 2018). These ice crystals then precipitate toward the polar surface, and hence lakes in which their chemical and potential astrobiological impacts are totally unknown.

Knowing the composition of Titan's aerosols is important to understanding the chemistry occurring in the neutral atmosphere and to better characterize their impact on the radiative budget. Their composition is directly linked to the spectral variation of their refractive index, while their morphology (size, shape) affects their absorption and scattering properties. Aerosol refractive indices were roughly determined from Cassini/VIMS and CIRS observations in the 100-500km range (Rannou et al. 2010; Vinatier et al. 2012) and haze optical properties were determined *in situ* at a single location below 150km near the equator from Huygens' measurements of their scattering properties (Tomasko et al. 2008a; Doose et al. 2016). Knowledge of haze optical properties is also of prime importance to: (i) retrieve the surface spectra through the atmosphere in $CH_4$ windows in which aerosols have spectral contribution, especially at 2.8 µm near the strong N-H absorption peak; (ii) to understand the cloud formation as aerosol composition and their morphology influence the gases condensation. Titan's aerosols are also probably quite representative of the haze that seems to be naturally produced in $CH_4$-rich atmospheres like those of Pluto, Triton, and probably many exoplanets.

Hence, many open questions cannot currently be answered without a dedicated mission to Titan:

- **[1] What generates Titan's atmospheric superrotation and what maintains it?**
- **[2] How do the polar vortices form, evolve, and dissipate?**
- **[3] What is the complexity of the chemistry attained inside the polar vortices? What are the composition and the structure of the massive stratospheric polar clouds?**

- **[4] What are the chemical composition, optical properties, and spatial distribution of aerosols in the main haze layer?**

*2.2.3 Proposed instrumentation and mission concept to address the open questions*

**Understanding the global dynamics of Titan's atmosphere (open question 1)** requires:

- **An orbited visible and near-IR imager,** monitoring in detail atmospheric wave activity around the equinox and when the polar vortex is forming. Monitoring the haze distribution using an orbited imager will be the only way to constrain Titan's atmospheric wave activity, as they are not detectable from Earth's orbit even from the largest facilities (JWST, ELT).
- **An orbited sub-mm heterodyne spectrometer** to measure the wind profiles (including the 80km altitude wind minimum) and probe the pole-to-pole structure along seasons. It will be possible to measure the meridional circulation speed for the first time (predicted to be at most of ≈1m/s) by integrating ≈10 minutes with this type of instrument, which has a precision of 3m/s in 1 minute integration (Lellouch et al. 2010).
- **An orbited radio occultation experiment** to probe the pole-to-pole structure of the 80km altitude wind minimum along seasons as performed by Flasar et al. (2013) from the very limited number of radio occultations performed by Cassini.
- **An *in situ* wind measurement experiment** during the descent of one/several landers to get the precise profile and the changing directions of the winds along the descent.

**Understanding the polar vortex structure and its evolution (open question 2)** requires an orbiter with a **sub-mm spectrometer** to directly measure the vortex zonal winds from a high inclination orbit (however a strictly polar orbit would not permit measurement of zonal winds at high latitudes, Lellouch et al. 2010). Their spatial/vertical structures with time will be determined for the first time from the lower stratosphere to 1000km. This will provide the first 3-D view of the vortex winds as well as its thermal structure and its composition in the middle and upper atmosphere where the vortex structure is totally unknown (as Cassini only probed altitudes below 600km from thermal infrared measurements).

**The chemistry inside polar vortices (open question 3), and aerosol composition and optical properties (open question 4)** will be revealed with the combination of:

- **An orbited sub-mm spectrometer** providing the mixing ratios of many photochemical species from the stratosphere up to 1000km.
- **An orbited UV spectrometer** whose observations of stellar/solar occultations will probe the aerosol vertical structure from the 400-1400km range as well as the density profiles of $CH_4$, HCN, $HC_3N$, and species that cannot be observed in the sub-mm spectral range: $N_2$ (above ≈1100km), $C_2H_2$, $C_2H_4$, $C_4H_2$, and $C_6H_6$.
- **An orbited visible and near-IR spectral-imager** is necessary to determine, from the reflected sunlight radiation, the optical constants of aerosols in the visible and near-IR spectral range, and how they vary in the atmosphere and with season, especially inside the Winter polar vortex in which enriched air coming from above can modify their composition. This instrument will also probe the vertical profile of the aerosol extinction coefficient and will reveal the structure and composition of the stratospheric icy clouds.
- **An orbited far- to mid-IR spectrometer is** necessary to determine the vertical and spatial variations of the aerosols' optical constants in the far- and mid-IR spectral range where they emit IR radiation, and if these properties seasonally and spatially vary in the atmosphere. It will also allow the determination of $C_2H_2$, $C_2H_6$, $C_2H_4$, $C_4H_2$, $C_6H_6$ vertical profiles below 500km, which cannot be observed in the sub-mm (no spectral lines) nor UV (sensitive to altitudes higher than 450km). This instrument should include the spectral range 1450 – 1900$cm^{-1}$ (7-5µm) that was not observed by Cassini. This range is of particular interest because it displays a strong peak in the haze extinction cross-section there, as detected from ISO/SWS observations (Courtin et al. 2016). This instrument will also be necessary to determine the ice composition of the stratospheric polar clouds (e.g. Anderson et al. 2018).

- **An *in situ*** high-resolution mass spectrometer (e.g. **CosmOrbitrap)** (on a lander/drone) will allow the detailed composition of the air, cloud, aerosols of the polar region through the descent.
- **An *in situ* imager/spectral radiometer**: to observe solar aureole during the descent, derive the column opacity, the average aerosol and cloud particle size and their spectrum.
- **An *in situ* nephelometer/particle counter:** to determine the aerosol and cloud particle densities and size distribution.

### 2.3 The lower atmosphere: clouds, weather, and methane cycle

#### *2.3.1 Current knowledge*

The deepest 80km of Titan's atmosphere contains the deep stratosphere and troposphere. This region, currently poorly known (known mostly from its cloud activity and the *in situ* measurements performed by Huygens), is of particular interest because the interaction between the surface and atmosphere occurs in the boundary layer (in the deepest 2km) through convection. Some strong convective events can occur sporadically with cloud tops reaching the tropopause at ≈40km altitude (Griffith et al. 2005). Aerosols also play an important role below 100km, as they serve as condensation nuclei, evolve under UV radiation and are removed by sedimentation and rainfall. The conditions of temperature and pressure on Titan allow the presence of a hydrological methane cycle in the lower atmosphere, very similar to the Earth's water cycle (Mitchell and Lora 2016; Hayes et al. 2018). The weak solar flux reaching Titan's surface, and the generally dry conditions in the lower troposphere, lead to relatively rare tropospheric clouds (Griffith et al. 2005, 2006; Rodriguez et al. 2009, 2011; Turtle et al. 2009, 2011a, 2018; see **Figure 5**). Despite the scarcity of the observed tropospheric clouds, Cassini revealed their diversity, including small patchy convective clouds, tropical storms associated with precipitation (Turtle et al. 2011a,c), and stratospheric polar clouds. Large tropospheric clouds are thought to be composed of large methane droplets in ascending motions while thinner high-altitude clouds are made of smaller ice crystals of photochemical by-products ($C_2H_6$, $C_2H_2$, HCN, and other nitriles and hydrocarbons) in descending air (Rannou et al. 2006; Griffith et al. 2005, 2006; Barth and Toon 2006; Barth and Rafkin 2010; Hueso and Sanchez-Lavega 2006; Lavvas et al. 2011c). The mixing ratios and vertical profiles of $CH_4$ and of photochemical byproducts are also poorly known and only a few species have been sampled *in situ* at one latitude by Huygens (Niemann et al. 2005, 2010). These ratios are important to constrain as they directly impact the formation of clouds and the $CH_4$ cycle in Titan's deep atmosphere.

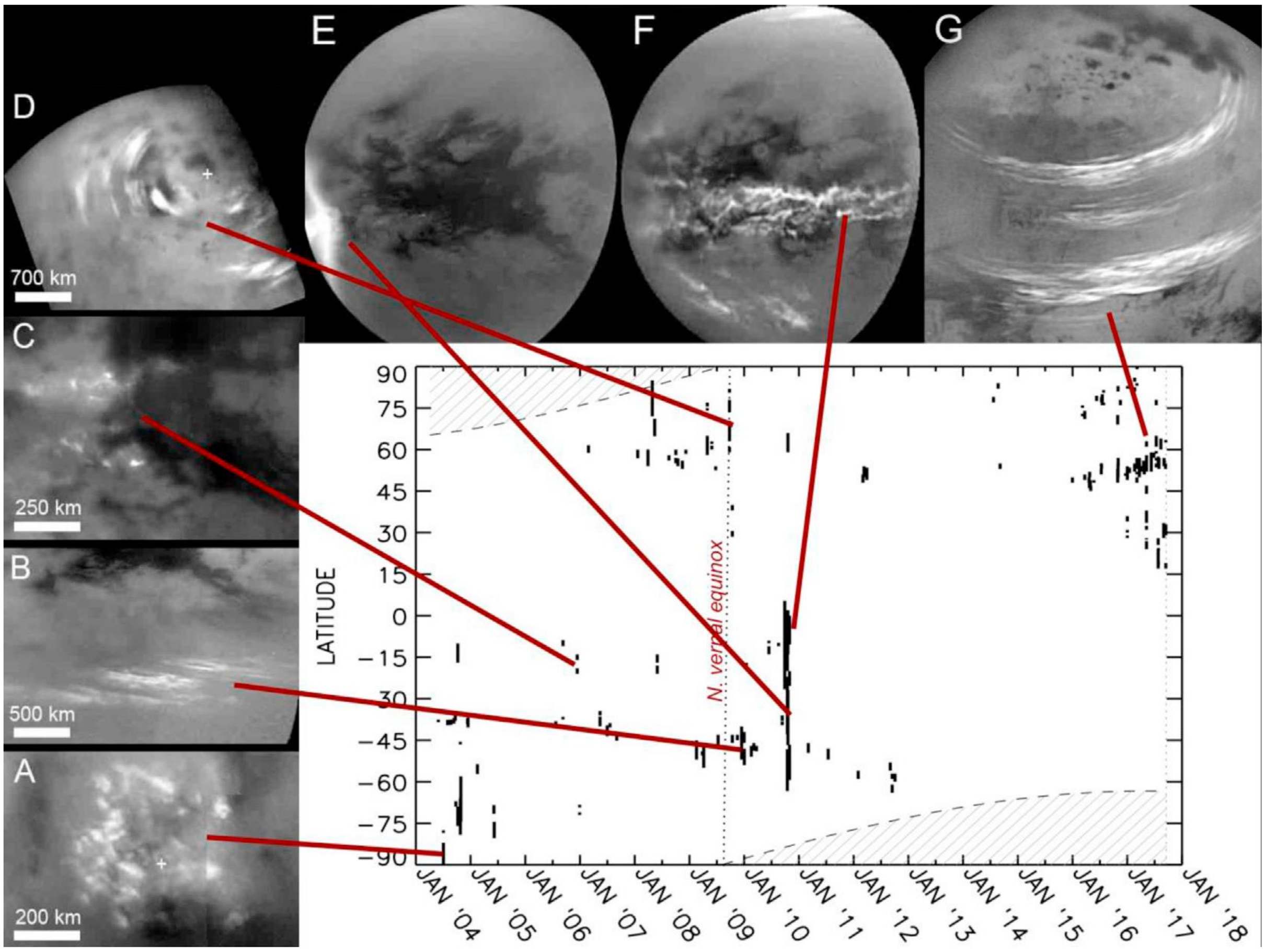

Figure 5: Cassini/ISS images of different types of clouds (A–G) and graph of the latitudes at which clouds have been observed over the mission, spanning May 2004–September 2017 (after Turtle et al. 2018).

### *2.3.2 Open questions*

$CH_4$, which plays a key role in the complex chemistry of Titan's atmosphere, comes from the surface and/or subsurface (**Section 3.5**). Monitoring the vertical and latitudinal distribution (Adamkovics et al. 2016; Lora and Adamkovics 2017) and seasonal evolution of tropospheric methane humidity may allow us to identify the $CH_4$ main evaporation sources (e.g. polar lakes, hypothetical tropical lakes, or ground humidity). Cassini/CIRS observations also revealed unexplained large latitudinal variations of the methane abundance in Titan's stratosphere (Lellouch et al. 2014), which could be the result of methane injection from strong tropospheric convective events.

Even if several mechanisms have been proposed to explain the diversity, localization, and seasonal evolution of observed tropospheric clouds, including planetary waves (Mitchell et al. 2011), global circulation (Mitchell et al. 2006; Rannou et al. 2006; Lora et al. 2015), topography, and boundary layer processes, a clear understanding of how clouds form, evolve, and dissipate is missing. Their composition and the size of the cloud droplets are highly linked to their formation mechanisms and the composition of the atmosphere, both being currently undercharacterized. Getting a more complete climatology of Titan's clouds will provide strong constraints on the atmospheric circulation, the methane transport, and the dominant mechanisms of cloud formation (global circulation, planetary waves, etc). A key question related to cloud formation and coverage is where and at which season it rains, and what are the precipitation rates. In particular, the dynamics, frequency, and precipitation rates of convective methane storms are of prime interest to explain the formation of fluvial valley networks and equatorial dunes (Mitchell et al. 2008; Charnay et al. 2015; Faulk et al. 2017), see **Sections 3.1 and 3.2**.

Even though different theoretical models of condensation profiles exist (Lavvas et al. 2011c; Barth 2017), a key question is to know the exact composition of aerosols during the condensation process and their evolution under long UV exposure. Another remaining question is to understand the chemical nature of these aerosols when they reach the surface and their potential evolution after sedimentation depending on the relative humidity at Titan's surface.

The only direct measurements that we have about the wind speed in Titan's lower troposphere come from the Huygens probe descent, at a single epoch and a single location near the equator. A key question is to know the direction and speed distribution of surface winds as well as their seasonal evolution. In particular, atmospheric models predict that mean surface winds should be westward in the equatorial region (as trade winds on Earth), while dunes propagate eastward (Radebaugh et al. 2008). The sand transport may be dominated by strong and rare eastward gusts produced by vertical mixing or by methane storms at the equinox (Tokano 2010; Charnay et al. 2015). In the same manner, large dust storms may have been detected by Cassini in the arid tropical regions of Titan (Rodriguez et al. 2018; Karkoschka et al. 2019), but the strength of the surface winds able to generate them are still unknown. Another remaining question is to understand the atmospheric circulation over Titan's lakes and the timing and frequency of their wave activity.

Answering these open questions regarding the physico-chemical properties of Titan's lower atmosphere remains a major goal for future missions to address. They are summarized below:

- **[1] What are the characteristics of the $CH_4$ cycle on Titan?**
- **[2] How do Titan's clouds form and evolve? What is their precipitation rate?**
- **[3] What is the wind regime near the surface?**
- **[4] What is the chemical composition of aerosols and how does it evolve through sedimentation and at the surface?**
- **[5] What are the chemical exchanges between the condensed ice/droplets and gas-phase atmosphere?**

*2.3.3 Proposed instrumentation and mission concept to address the open questions*

**Understanding the methane cycle in the troposphere and its injection in the deep stratosphere (open question 1)** requires:

- **An orbited far and mid-IR spectrometer** and **sub-mm spectromete**r to derive the $CH_4$ abundance, independently of the temperature in the lower stratosphere, all over the globe and especially above the northern lake region.
- **An *in situ* near-IR spectrometer** on a drone and/or a/several lander(s) to probe the methane abundance (humidity) below the tropopause and near the surface.

**Understanding cloud formation and humidity conditions in the troposphere requires (open question 2)**:

- **An orbited radio occultation experiment** to determine the temperature profile from 150km down to the surface.
- **An orbited visible and near-IR camera** to monitor cloud activity and precipitation signatures on the surface.
- **An orbited near-IR spectrometer** to derive the cloud composition and information on the $CH_4$ droplet size as well as other crystal sizes ($CH_4$, $C_2H_6$).
- **An *in situ* nephelometer/particle counter, camera/spectral radiometer, and high-resolution mass spectrometer (e.g. CosmOrbitrap)** on a drone or on one/several lander(s) (with data acquisition during the descent) to determine the chemical atmospheric composition and species vertical profiles, the aerosol and cloud particles composition, their size distribution, and the cloud droplet phase (liquid, crystalline), and to look at the solar aureole to measure the scattered light and transmission through the atmosphere.

**Understanding the wind regime at/near the surface (open question 3)** requires ***in situ* anemometer** on a drone and/or on one/several landers localized in the northern lake region and possibly at the equator.

**Understanding aerosol compositions and how they evolve at the surface (open questions 4 and 5)** requires a high-resolution mass spectrometer (e.g. **CosmOrbitrap)** (on a drone) and **a far- to mid-IR spectrometer** and **a UV spectrometer** to determine the composition of the aerosols when they reach the surface.

Some of the above-mentioned *in situ* instruments (near-IR spectral imaging capability, anemometer) will be carried by the Dragonfly mission (Lorenz et al. 2017) that will study the dune regions and the Selk impact crater with a planned arrival by 2034, i.e. during an expected "quiet" period, as the northern Spring equinox will occur in 2039. A particular objective of our future Titan mission is to monitor the seasonal changes around equinoxes, and thus, it will be highly complementary to the Dragonfly mission. The best science return from the two missions would be obtained if timing overlap could occur. Note that the required measurements to answer these open questions cannot be achieved from terrestrial ground-based or space-borne facilities.

## 3. Science Goal B: Titan's geology

Titan has diverse and strikingly familiar landscapes (mountains, rivers, seas, lakes, dunes, impacts, etc), see **Figure 6**. Many of those Titan surface morphologies are presumed to originate from exogenic processes (Moore and Pappalardo 2011), involving a complex and exotic climatology based primarily on the methane cycle, analogous to the hydrological cycle on Earth (Atreya et al. 2006). Despite the heavy exogenic influence, endogenic processes (including a possible past and/or present tectonic and cryovolcanic activity) may also be at play. Despite the recent and remarkable progress accomplished so far in Titan's knowledge thanks to the successful and long-lived Cassini-Huygens mission, there are numerous key issues regarding Titan's geological history, and its link to Titan's climate history, that remain poorly constrained.

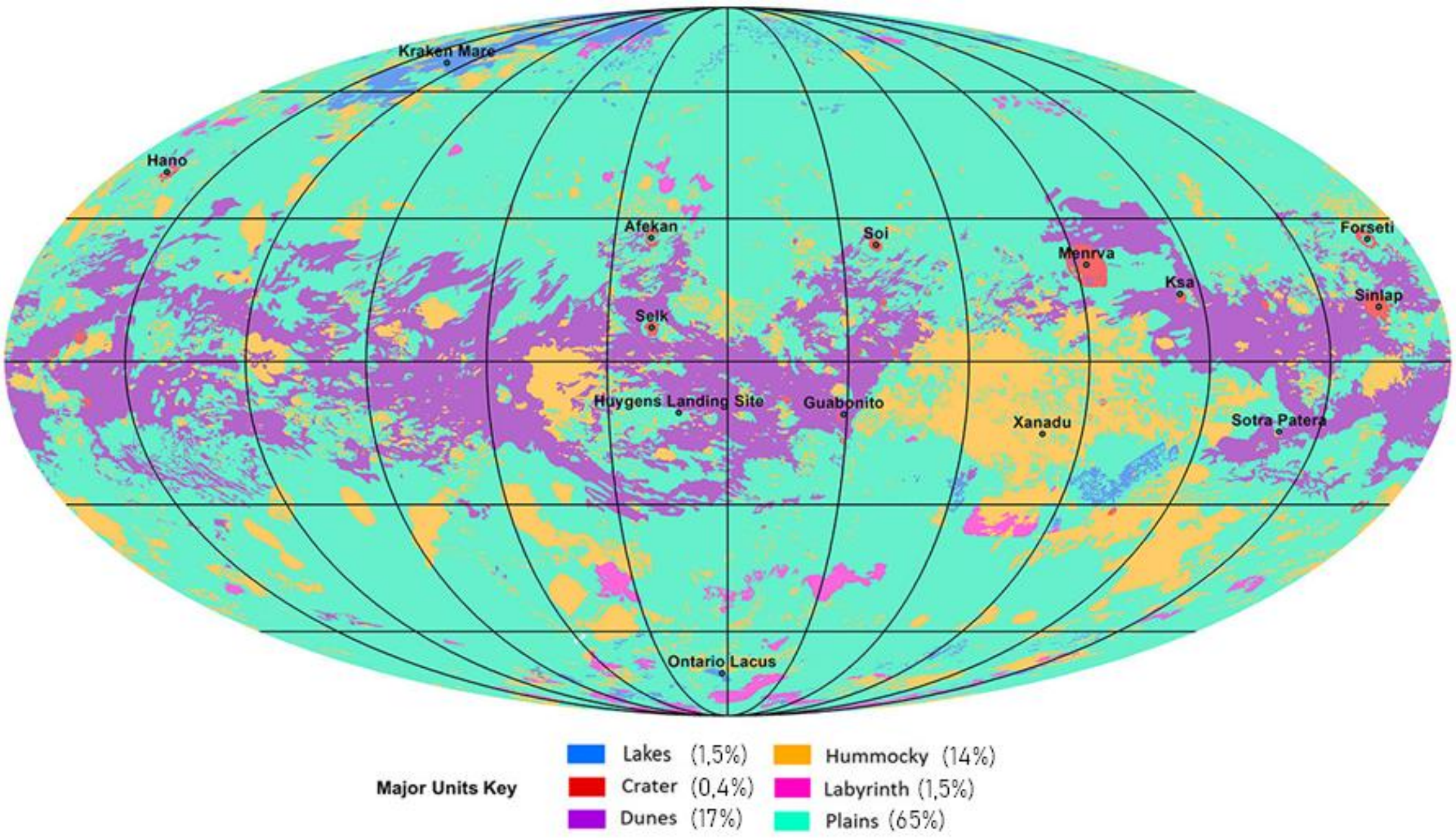

Figure 6: Global map of Titan's major geomorphological units (after Lopes et al. 2020).

## 3.1 Aeolian features and processes

### *3.1.1 Current knowledge*

Cassini observations revealed that dunes are Titan's dominant aeolian landform (**Figure 6**). The latitudinal distribution of these features is indicative of the different types of climates that Titan experiences or has experienced in the past. Dunes, in particular, provide a powerful tool to investigate the sedimentary and climatic history of the arid and/or semi-arid environments likely to prevail at Titan's tropics.

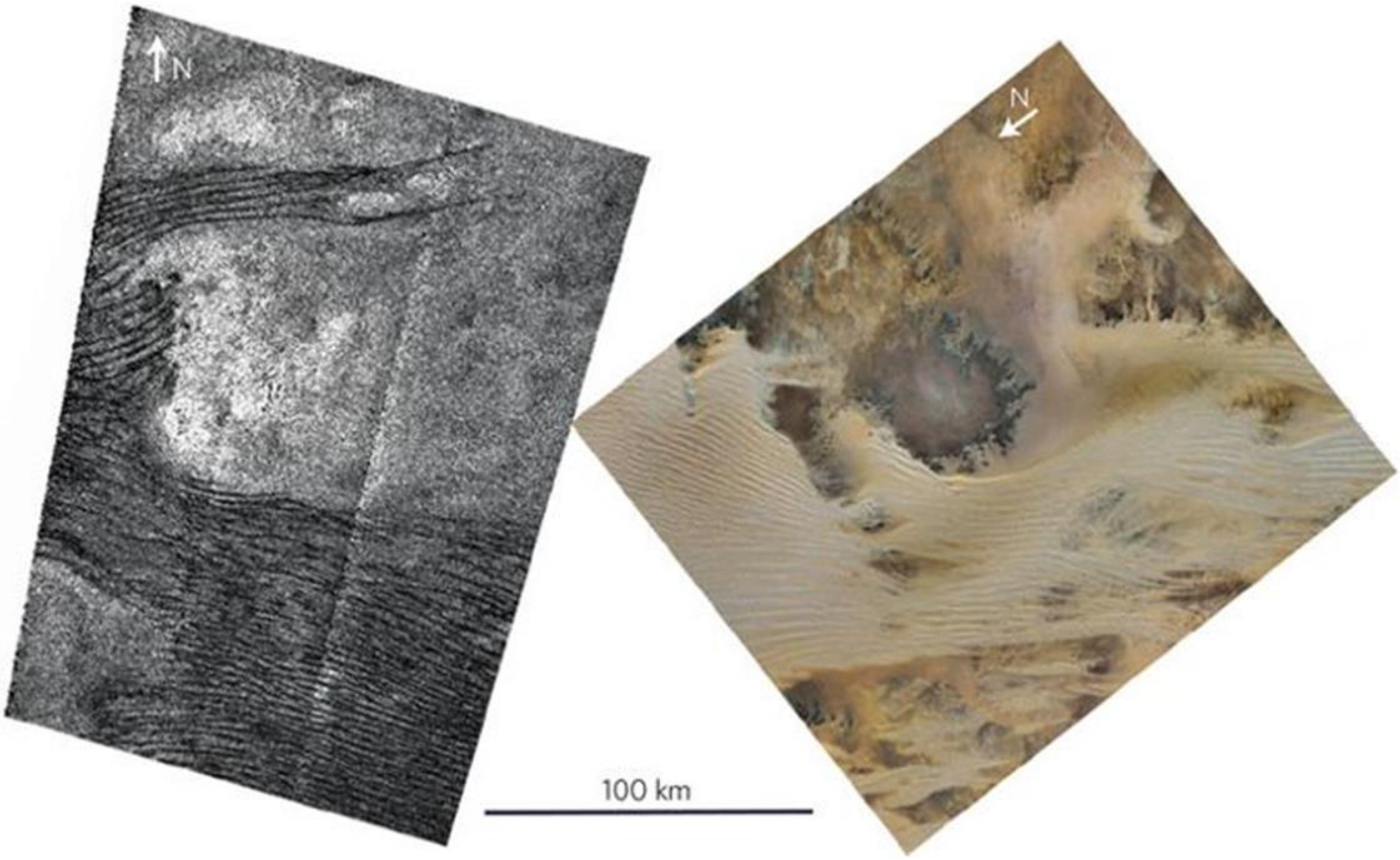

Figure 7: Titan's linear dunes (left), as seen by the RADAR/SAR on board Cassini. Interaction between dunes and topographic obstacles on Titan (left) and in Libya (right) (from Radebaugh, 2013). Right image courtesy of Google Earth.

In Cassini images, Titan's dunes appear as long, narrow, and RADAR-dark features as opposed to RADAR-brighter substrate, presumably because sand dunes are absorbent and smooth at the 2.17 cm wavelength (**Figure 7**). The dunes are generally 1–2km wide, spaced by 1–4km and can be over 100 km long (Lorenz et al. 2006; Radebaugh et al. 2008). Limited estimation of heights from radarclinometry and altimetry suggests they are 60–120m height (Neish et al. 2010; Mastrogiuseppe et al. 2014).

Merging all Cassini's RADAR and infrared observations, despite incomplete or at low average resolution at numerous places, Rodriguez et al. (2014) and Brossier et al. (2018) estimated by extrapolation the geographic distribution of Titan's dunes at global scale. Titan's dunes cover up to ≈ 17% of the moon's total surface area (≈ $14x10^6 km^2$, 1.5 times the surface area of the Sahara desert on Earth), the same as early estimates from Lopes et al. (2010). 99.6% of the imaged dunes are found within ±30° latitude forming an almost complete circum-Titan belt, with the notable – unexplained – exception of the Xanadu region. The extent of the dunes indicates that sands have been generated on Titan in great volumes and transported by wind, and that processes have acted on the surface long enough to produce extensive and morphologically consistent landforms (Radebaugh 2013).

Cassini observations indicate that Titan's dunes interact with topographic obstacles in a way that indicates general West-East transport of sand (e.g. Radebaugh 2013), see **Figure 7**. Within the current uncertainties, their size, morphology, and relationship with underlying terrain, and their style of collection are similar to large, linear dunes in Earth's sand seas of the Sahara, Arabia, and Namibia (Lorenz et al. 2006; Radebaugh et al. 2008; Le Gall et al. 2011; 2012). Such dunes on Earth typically form under a bimodal wind regime (Fryberger and Dean 1979; Tsoar 1983). A recent model calls on a dominant, slightly off-axis wind and a secondary wind causing sand flux down the dune long axis (Courrech du Pont et al. 2014; Lucas et al. 2014a). A fundamental challenge raised by the RADAR observations of the dunes is the eastward direction of sand transport (Lorenz et al. 2006; Radebaugh et al. 2010). This contrasts with expectations from climate models that low-latitude, near-surface winds should generally blow to the west. A possible solution appears to be that the dunes reflect strong but infrequent eastward winds, either associated with vertical mixing in the atmosphere at equinox leading to strong westerly gusts (Tokano 2010) or methane rainstorms having a similar effect (Charnay et al. 2015). Note that equatorial methane rainstorms may be associated with regional dust storms, possibly observed by Cassini (Rodriguez et al. 2018). Additionally, convergence of the meridional transport predicted in climate models can further explain why Titan's dunes are confined within ±30° latitudes, where sediment fluxes converge (Lucas et al. 2014a; Malaska et al. 2016).

Titan's dunes are not only consistently dark to Cassini's RADAR but they are also some of the infrared-darkest materials seen by the Cassini/ISS cameras (Porco et al. 2005; Karkoschka et al. 2017), and have a low albedo and red slope in the near-infrared as seen by VIMS (Soderblom et al. 2007; Barnes et al. 2008; Clark et al. 2010; Rodriguez et al. 2014). This strongly indicates that the dunes are smooth, homogeneous, and primarily dominated by organic sand, presumably settling from the atmosphere and further processed at the surface (Soderblom et al. 2007; Barnes et al. 2008; Le Gall et al. 2011; Rodriguez et al. 2014; Bonnefoy et al. 2016; Brossier et al. 2018), making the dunes one of the largest carbon reservoirs at Titan's surface (Lorenz et al. 2008a; Rodriguez et al. 2014). Bonnefoy et al. (2016) extracted distinct dune and interdune spectra and emissivities from most of Titan's major dune fields. Their results indicate sand-free interdune areas of varying composition, implying that the sand dunes have been active on recent geologic timescales.

In addition to the dunes, other aeolian features and landforms on Titan's surface may have been identified by Cassini. These are possibly wind streaks and yardangs, or wind-carved ridges. The wind streaks are visible in ISS images as bright features that extend in the downwind direction from obstacles (e.g. Porco et al. 2005; Lorenz et al. 2006; Malaska et al. 2016). They can be several tens of kilometers wide and long, can have flow-like, teardrop shapes, and appear as though wind has shaped the bright landscapes, and has deposited dark materials, most likely sand, in the low regions downwind of the obstacles. These features help indicate the direction of the winds, which also broadly parallels the linear dunes seen in Cassini RADAR images (Malaska et al. 2016). The mid-latitudes contain large, radar bright areas with parallel long lineations ≈1km wide, spaced by a few km, and tens of kilometers long that are interpreted to be yardangs (Paillou et al. 2014; 2016; Malaska et al., 2016; Northrup et al. 2018). These appear to have formed in easily eroded materials, similar to yardangs on Earth and Mars and further indicate the action of wind at moderate to high latitudes now or in the past (Northrup et al. 2018).

### *3.1.2 Open questions*

The complex interplay between the hydrocarbon cycle, atmospheric dynamics, and surface processes leading to the formation and dynamics of Titan's aeolian features is still far from being fully understood. The precise composition, grain size, and mechanical properties of Titan's sediment, its total volume, sources, transport dynamics, and pathways at global scale still require further investigation.

Important open questions include:

- **What is the precise – not extrapolated – geographic distribution of Titan's aeolian landforms?**
- **What is the precise morphometry of the dunes, including length, width, spacing, and height and does it vary across Titan?**
- **What is the wind regime responsible for dunes' and other aeolian landforms' morphology and orientation?**
- **Are dunes and other aeolian landforms still active today? Are there changes over observable timescales?**
- **What are the sources and sinks of Titan's sand? Can we ascertain the pathways of sediment transport? Why are there no dunes in the Xanadu region?**
- **What is the composition, grain size, degree of cohesion and durability of the dune material?**

*3.1.3 Proposed key instrumentation and mission concept to address those questions*

Most of the open questions related to Titan's global distribution and properties of Titan's aeolian landforms (statistics on dunes' width, spacing, shape, and height) can be addressed by a **Titan orbiter**, instrumented with a **multi-wavelength remote sensing package** (e.g. near-infrared in Titan's atmospheric windows' spectral range and microwaves) providing decameter-scale spatial resolution and complete coverage by the end of the mission. This can be achieved either using a **SAR system** or **an infrared camera** working in one of the methane spectral windows. An **imager at 2µm** would provide the best tradeoff between signal to noise, atmospheric transparency, and low aerosol scattering effects. The study of global sediment pathways, sources, and sinks would highly benefit from the measurement of global altimetry. A **global "high-resolution" topography** is really the fundamental missing piece from Cassini.

In order to study the accurate morphometry of a selection of aeolian landforms and the physico-chemical properties of Titan's sediment at the grain scale, a **mobile *in situ* probe** with **surface sampling** **and analysis** (e.g. a CosmOrbitrap-based high-resolution chemical analyzer), **imaging,** and **spectral capabilities** is required.

## 3.2 Fluvial features and processes

### *3.2.1 Current knowledge*

One of the most striking observations after Cassini-Huygens is the channel networks (Collins 2005; Lorenz et al. 2008b; Lunine et al. 2008; Burr et al. 2006, 2009; Black et al. 2012; Burr et al. 2013), see **Figure 8**. Similar to Earth, Titan may experience complex climate-topography-geology interactions, involving surface runoff and subsurface flows.

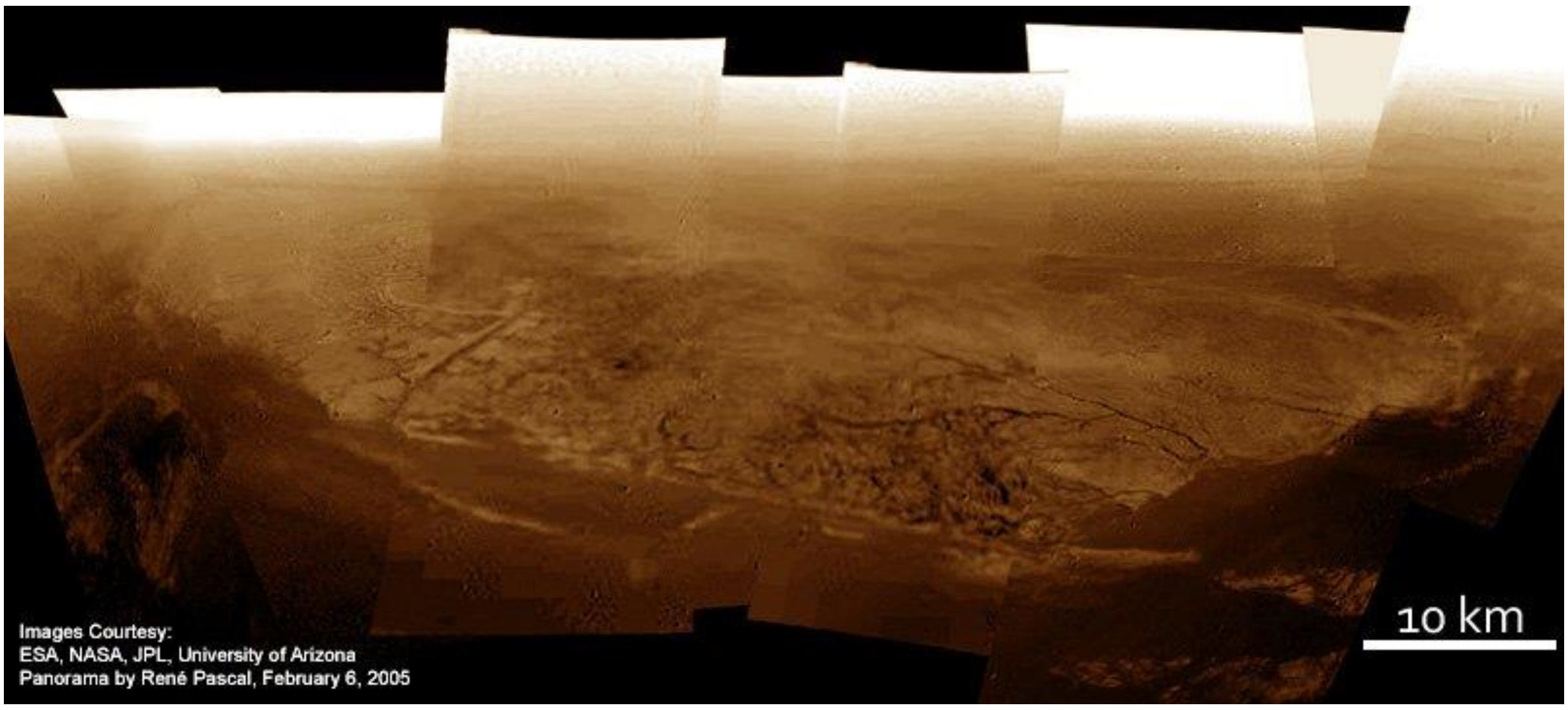

Figure 8: Panorama reconstituted and colored from images taken by Huygens' DISR instrument when the descent module was between 18km and 6km altitude. One can clearly see a hill cut by numerous channels at the foot of which seems to have been a darker plain. Huygens landed on this plain a few moments later. (credits: ESA/NASA/JPL/University of Arizona/René Pascal)

Valley networks are distributed at all latitudes (Lorenz et al. 2008b; Burr et al. 2009; Lopes et al. 2010; Langhans et al., 2012) and have a wide variety of morphologies: canyons and highly dissected plateaus at the poles (Poggiali et al. 2016; Malaska et al., 2020), dendritic and rectilinear networks globally distributed (Burr et al. 2013), meandering features in the south polar region (Malaska et al. 2011; Birch et al. 2018), and braided rivers at the equator (Lucas et al., 2014b). Near the poles, most of the fluvial networks are connected to empty or filled lacustrine features (cf. **Section 3.3**). The presence of canyons implies stratified bedrock with an alternation of weak and strong layers. Rectilinear network channels rather suggest fractured bedrock controlling their geometry. As for meanders and braided rivers, they are mainly controlled by the stream slope and the sediment load. Due to the hectometer-scale resolution of the Cassini RADAR, we have been constrained to studying only the largest valley networks on Titan. We therefore have a limited idea about the extent to which Titan's landscapes are dissected by fluvial networks. The one exception to this is the region where the Huygens lander descended, where descent images, with a decameter spatial resolution (Tomasko et al. 2005), showed a highly dissected network of dendritic valleys (Perron et al. 2006), indicating that river networks at that scale may be far more frequent on Titan than what is only inferred from the lowest resolution observations (**Figure 8**).

The mere presence of channelized flow conduits implies that the surface material can be eroded either physically or through dissolution, and that flows of sufficient magnitude, either from precipitation or groundwater, are able to erode Titan's surface. However, using estimates for the initial topography and erodibility of the substrate, channels may be very inefficient agents of erosion on Titan (Black et al. 2012) or there may be a gravel lag deposit that inhibits erosion under Titan's current climate (Howard et al. 2016).

In some locations, fluvial channels terminate in alluvial or fluvial fans, distributary landforms that indicate a transition from a high to a low elevation (Radebaugh et al. 2016; Birch et al. 2017). These are fairly low in slope, and in some cases can run out to large distances, indicating the carrying power by methane fluid of sedimentary rock (Radebaugh et al. 2016). At the end of the Cassini mission, these landforms are supposed to be rare and randomly distributed across the surface (Birch et al. 2017). This may indicate there is not frequent rainfall that can generate surface erosion, or that topographic gradients are gentle on a global scale such that these landforms are not readily generated due to a hypothetically high incision threshold. Again, the coverage and spatial resolution of Cassini observations are not sufficient and Titan's topography is not sufficiently well known, to definitively conclude on that question.

*3.2.2 Open questions*

Critical unknowns remain following the Cassini mission. Better estimates for the physical and chemical properties of both the bedrock and the fluid (including frequency and magnitude of rainfalls vs. latitude) are needed to provide better understanding about the role that fluvial channels have played in sculpting Titan's surface (Burr et al. 2006; Cordier et al. 2017; Malaska et al. 2017; Richardson et al. 2018). Due to the limited coverage and spatial resolution of the Cassini remote sensing package, we have only limited constraints on the fluvial channel geographic distribution and morphologies. What is the latitudinal (hence climatic) forcing on the dominant mechanism of Titan's river network formation and evolution (which of them are alluvial channels formed in sediments or rather incisional channels) are still open questions. For instance, we still do not know if there are any systematic variations in morphologic type that may be indicative of bedrock heterogeneities (Burr et al. 2013) and/or variations in flow conditions (stream velocity, stream gradient, sediment load) and climate forcing (Moore et al. 2014). Also, the possibility that Titan is dissected everywhere at the scale observed by Huygens, as it is the case on Earth, implying that Titan's landscape may be dominated by hillslopes, is still not confirmed. As hillslope processes control the sediment supply to rivers, it is important to clarify if slopes are made of consolidated material, or if they are covered with loose sediment (i.e., granular media). In the first case, landslides will erode the bedrock, in the latter slope creep will shape the landscape preferentially.

Many questions cannot be answered by the analysis of data from Cassini-Huygens or telescopes:

- **What is the complete geographic distribution of river networks down to the decametric scale?**

- **How do river network morphology types vary with location?**
- **What are the processes at play forming the rivers (incision and/or dissolution) and what is the nature of the eroded material?**
- **What is the frequency at which runoff occurs and river channels are filled?**
- **What are the ages and the current activity of the fluvial channels? How does this activity vary with latitude?**

*3.2.3 Proposed key instrumentation and mission concept to address those questions*

Answers to these questions require observations with a resolution finer than the scale of fluvial dissection (10's of meters). A **long-lived Titan orbiter** with a near-polar orbit and a **high-resolution remote sensing package** (down to decameter) will provide both the global coverage and needed repetitiveness (1) to build a consistent global map of the fluvial networks' distribution, (2) to provide a deeper look into their precise morphologies, and possibly (3) to build digital elevation models of a variety of river networks by photogrammetry and/or radargrammetry. The measurement of altimetry at the global scale would provide a **global "high-resolution" topography map** that is missing from Cassini and is essential for any geological studies (morphology, erosion…) of Titan. **Spectral capabilities are needed at both decameter spatial and high spectral (R>1000) resolutions** (**Figure 11b**) in order to help constrain the composition and texture of the eroded material, bedrock, and transported sediment. The spectral identification of the surface component will always be limited by the strong atmospheric absorption unless we develop a spectral-imager with a very high spectral resolution within the broader diagnostic methane windows (such as the 2µm, 2.7µm and 5µm windows)

A **mobile *in situ* probe** may be of great help to provide an unprecedented detailed view of the morphology of the river networks, the shape and size of the sediment, and the composition of the involved materials (fluid, sediment, substrate, and bedrock) with **remote sensing instruments** and **sampling capabilities** (including a CosmOrbitrap-based high-resolution chemical analyzer). The flexibility of an autonomous aerial drone would in addition provide the possibility to realize super high-resolution digital elevation models (down to the centimeter), allowing the analysis of river dynamics down to the scale of boulders and pebbles. This could be done at high northern latitudes to complement the similar measurements that will be performed in the equatorial region by the Dragonfly octocopter.

### 3.3 Seas and lacustrine features and processes

Titan's surface conditions (1.5 bar, 90 - 95 K) are close to the triple point of methane (and in the liquid stability zone of ethane), which allows standing liquid bodies to exist at the surface (Lunine and Atreya 2008). During the Cassini mission, Titan's surface has been unevenly mapped by all the imaging instruments (RADAR in SAR mode, VIMS imaging spectrometer, and ISS imaging subsystem) with various spatial resolutions (a few kilometers for ISS and VIMS, a few hundreds of meters for the RADAR), extent (global coverage with ISS and VIMS, 50 % of the surface with the RADAR at best 1500 m pixel resolution), and wavelengths (infrared and microwaves), looking for signs of these liquid bodies. Lacustrine features (lakes and topographic depressions) were first observed in 2004 in the infrared with the Cassini/ISS observation of Ontario Lacus (Porco et al. 2005), the largest lacustrine depression in Titan's south polar region (**Figure 9**). Titan's large seas and lacustrine features were then observed in 2006 when flying over the northern polar regions (Lopes et al. 2007; Stofan et al. 2007), see **Figure 9**, as well as numerous fluvial features connected to the seas (cf. **Section 3.2**). Most of the liquids are currently located in the North, presumably as a result of orbital forcing (Aharonson et al. 2009; Hayes 2016).

#### *3.3.1 Current knowledge*

The distinction between Titan's seas and lakes is based on their respective morphologies and size. Titan's large seas (Kraken Mare, Ligeia Mare, Punga Mare) are several 100s-of-km-wide features with complex dissected shorelines shaped by rivers and drowned valleys from the nearby reliefs. While lacustrine features, regardless of their liquid-filling state, appear as closed rounded to irregularly lobate depressions, 10s to a few 100s-of-km-wide, usually organized in clusters lying in flat areas (Hayes et al. 2017).

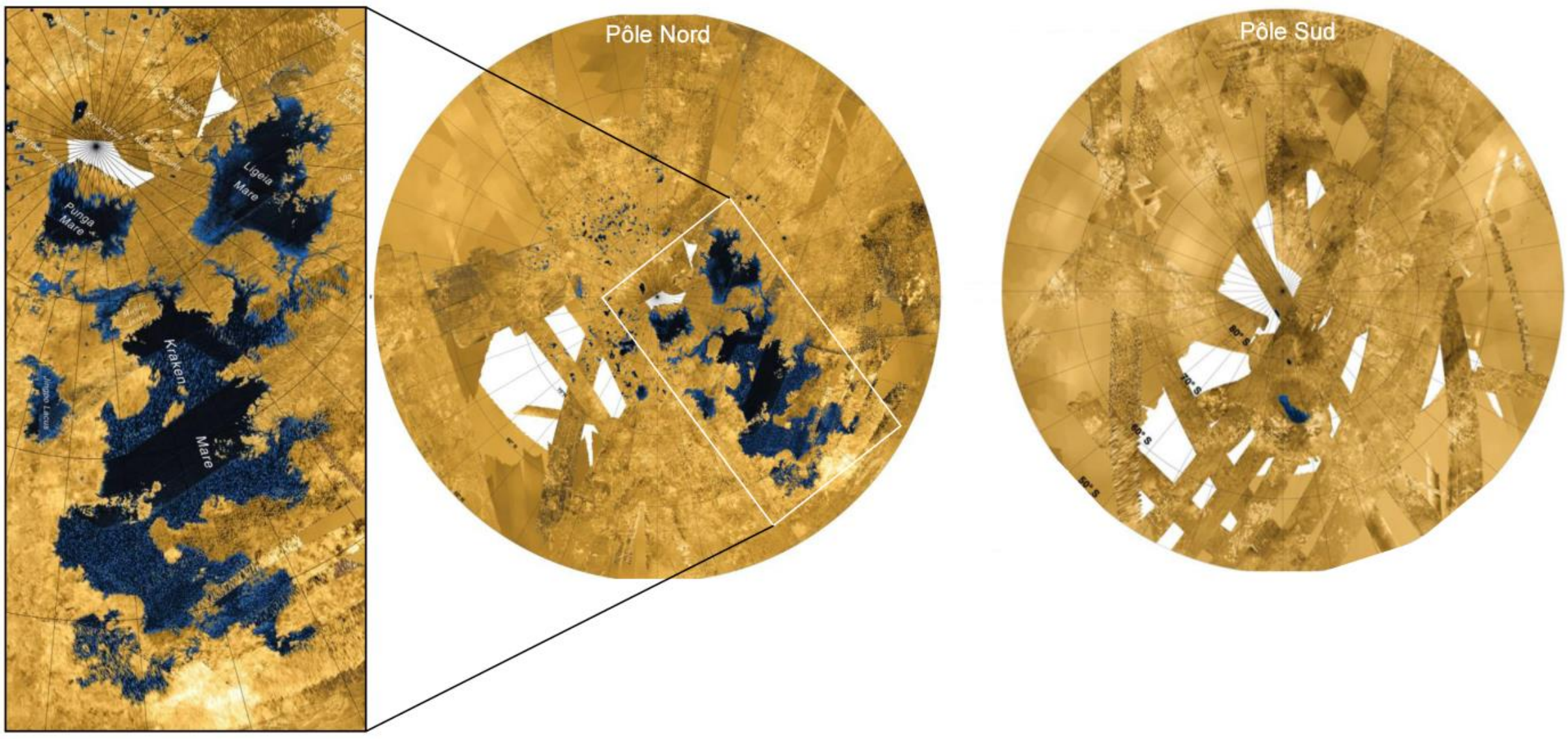

Figure 9: Seas and lakes of the North and South Poles of Titan. These maps are made from mosaics of RADAR images from Cassini, presented here in false color. The many dark areas, with very weak radar echoes (in blue), are lakes and seas of liquid methane and ethane. Inset: this region includes the three largest Titan Seas (Kraken Mare, Ligeia Mare and Punga Mare). (Credits: NASA/JPL-Caltech)

Cassini RADAR altimetry profiles, radargrammetry, or SARTopo techniques provided altitude estimates (at km-scale) of lacustrine depression depths ranging from 100 to 600 m, with an average depth of 200 +/- 100 m (Hayes et al. 2017). The depressions sometimes display 100 m-scale raised rims and ramparts, which could be some of Titan's youngest units (Solomonidou et al. 2020b). In a few cases, bathymetric profiles and rough composition could be determined by looking at double peaks in altimetry echoes over liquid-covered depressions. Thus, Ontario Lacus would be ≈50 m-deep in its lowest part (Mastrogiuseppe et al. 2018), while smaller lakes observed in the North would have depths varying between 20-30 m in some cases to more than 150 m in others (Mastrogiuseppe et al. 2019). Analyses of RADAR data provided a first estimate of the exposed volumes of liquids contained in the largest lakes and seas of Titan (70,000km$^3$; Hayes et al. 2018), and show that lakes and seas consist of a relatively small methane reservoir as compared to the atmospheric humidity, with varied compositions with respect to their locations and altitudes. Interestingly, Hayes et al. (2017) studied the elevation of the lacustrine depressions in the northern hemisphere and showed that liquid-filled depressions at regional scale are at similar altitudes and are systematically lower than empty depressions, both being located at higher altitudes than sea level. This suggests the existence of a subsurface connection between the lakes such as an alkanofer surface replenishing low-altitude depressions (Cornet et al. 2012a; Hayes et al. 2017). An unknown portion of the liquid hydrocarbons could therefore well be stored in a permeable subsurface (as also suggested by models aiming at reproducing the stable polar liquids (Horvath et al. 2016)), and not taken into account in the organic inventory calculation on Titan. The accurate estimate of the liquid composition, topography, and bathymetry of lacustrine features has a direct impact on constraining the amount of material removed and the total volume of liquid hydrocarbons stored in the seas and lakes.

Lakes and seas strongly differ in shape (Dhingra et al. 2018). The absence of well-developed fluvial networks at the 300m-scale of the RADAR/SAR associated with the lacustrine features, in addition to the fact that they seem to grow by coalescence in areas hydraulically and topographically disconnected from the seas, suggests that a distinct scarp retreat process is responsible for the formation and evolution of the lacustrine features on Titan (Cornet et al. 2012a; 2015; Hayes et al. 2016, 2017; Solomonidou et al. 2020b), but do allow for a volcanic collapse or explosion origin (Mitri et al. 2019; Wood and Radebaugh 2020). Among the hypotheses elaborated on to explain the formation of the lacustrine features, the thermodynamical, geological, and chemical contexts seem to favor the formation by karstic dissolution/evaporitic processes involving chemical dissolution/crystallization of solutes (soluble molecules) in solvents (in response to the rise or lowering of ground liquids; Cornet et al., 2012a, 2015; Cordier et al., 2016).

Depending on its exact composition, the liquid phase is more or less stable under Titan's surface conditions, ethane and nitrogen giving more stability to the liquids (Luspay-Kuti et al. 2012, 2015). While the current search for lake changes within the Cassini dataset is still debated (Turtle et al. 2011b; Cornet et al. 2012b; MacKenzie et al. 2019), Titan's surface exhibits hints of surface liquid changes at geological timescales. The identification of geological features that may have held liquids in the recent or distant past can help address this question. Currently, Titan possesses a climate which allows for the long-term accumulation of polar liquids and which bring liquids to the low latitudes only during torrential and sporadic events (Turtle et al. 2011c). Nonetheless, polar liquids are currently often most seen in the North as a result of Saturn's current orbital configuration (Aharonson et al. 2009; Hayes 2016; Birch et al. 2018), while the South of Titan exhibits large catchment basins (Birch et al. 2018; Dinghra et al. 2018). Their total area is equivalent to that of the northern seas (Birch et al. 2018). A few lacustrine features may have been detected at lower latitudes (Griffith et al. 2012; Vixie et al. 2015), which indicates that the climate has evolved through time, potentially reversing the preferred location of liquid accumulation to the South in the past. By constraining the past climate of Titan, looking at the superficial record of the changes at the surface, one can reconstruct a climate history that will take part in constraining the methane cycle on Titan.

*3.3.2 Open questions*

Despite many observations spreading over the 13 years of the Cassini mission (from Winter to Summer in the northern hemisphere), a number of open questions remain regarding the methane cycle and the evolution of Titan's seas and lakes. In particular, we still do not have a clear understanding of the formation scenarios of Titan's lakes, the precise composition of the liquid and substrate, the connectivity of the lakes and lake basins, the history and timescales of filling and emptying of lake liquids, the total volume of organics stored in the seas and lakes (and in a potential alkanofer), and how these organics are redistributed over seasonal to geological timescales.

Uncertainties remain on the shapes of these features in a three-dimensional sense due to the scarcity and accuracy of available topography data (currently provided at global scale and poor horizontal resolution by Corlies et al. 2017). Also, to date, only a few bathymetry profiles have been derived from RADAR altimetry data crossing liquid bodies. The bathymetry of Titan's largest sea, Kraken Mare, remains unknown. In the same manner, the exact composition of the solutes implied in dissolution/crystallization processes and of the solvent is to be determined. Radar altimetry data suggests the northern seas are primarily methane (Mastrogiuseppe et al. 2014), though VIMS observations also detected ethane, but in the southern lake Ontario (Brown et al. 2008). The mechanical properties of the substrate, influencing the landscape evolution in response to mechanical/fluvial erosion and hillslope processes that can also contribute to some extent to the surface evolution, are also to be determined. The way and timescale on which solids can accumulate over the surface to build the chemically eroded landscapes has also to be constrained, notably by characterizing the thickness of the organic sedimentary layer being eroded.

The remaining major open questions on lacustrine features and processes on Titan are the following:

- **What are the three-dimensional shapes of the lacustrine features in the polar regions?**
- **What is the true distribution of sub-kilometer lakes and what does this tell us about lake formation?**
- **How much liquid is stored in the lake and sea depressions, how do they connect with a subsurface liquid hydrocarbon table, and what is the true total inventory of organics in the polar areas?**
- **What are the exact compositions of the lakes and seas, how and why do they differ?**
- **By which geological processes do the lacustrine depressions and raised ramparts or rims form?**
- **What changes occur at the lakes and seas over seasonal/short timescales?**

*3.3.3 Proposed key instrumentation and mission concept to address those questions*

A global topographic map at high vertical (10's of meters) and horizontal resolutions (a few hectometers) is required to address the major open questions regarding the total surface and sub-surface liquid organic inventory on Titan. At regional scale, at higher resolution, it will also help to constrain the formation of Titan's seas and lakes and to connect their distribution and properties with the present and past climatic conditions.

A **long-lived Titan orbiter** with a near-polar orbit will be required, including a **SAR system**, a **Ground Penetrating Radar system** and/or a **high-precision altimeter**. An ***in situ*** **mobile/floating/submarine probe**, including a **spectral imager**, **electrical environment and meteorological packages**, and **sampling capabilities** (e.g. a CosmOrbitrap-based high-resolution chemical analyzer) would provide fundamental support to the questions of lakes and substrates' topography and composition, and local geologic and climatic conditions.

## 3.4 Impact craters and Mountains

*3.4.1 Current knowledge*

Cassini is the spacecraft that unveiled Titan's impact crater paucity (**Figure 6**), since prior to the Cassini era the impact cratering history of Titan was unknown and speculated to be similar to the other Saturnian moons. Hence, based on that fact, several hundred craters were expected to be found on the surface. However, Cassini observations showed that the resurfacing rate of Titan, due to its very active atmosphere – similar to Earth's – has modified and erased the majority of impact craters that formed in the past, especially in the polar regions, leaving only approximately 60 potential ones to account as craters (Wood et al. 2010; Buratti et al. 2012; Neish and Lorenz 2012). This scarcity of impact features indicates that the surface is geologically young, possibly about 500 million to a billion years (Tobie et al. 2006; Neish and Lorenz 2012). This impact crater identification was made after the analysis of Cassini RADAR images and VIMS infrared spectro-images. From these 60 potential impact craters, only 23 are labeled as "certain" or "nearly certain" (Lopes et al. 2019; Werynski et al. 2019).

The analysis of Cassini data showed that Titan's craters are subject to extensive modification due to the erosional activity including overlapping from fluvial channels, and infilling by sand (e.g. Le Mouélic et al. 2008; Soderblom et al. 2010; Neish et al. 2015; Brossier et al. 2018). In addition, the craters are not uniformly distributed across the surface: the Xanadu Regio area has the largest crater population and at the poles there is complete crater absence (e.g. Neish and Lorenz 2014). There are a number of theories to explain this anomaly in distribution including crater burial due to heavy methane deposition or crater degradation due to fluvial erosion near the poles (Moore et al. 2014; Neish et al. 2016). Nevertheless, this remains an open question. An additional mystery about Titan's impact craters is their chemical composition. There have been suggestions from analyzing crater with Cassini VIMS infrared data, that the very top layer of the impact craters seems to be dominated by atmospheric tholin-like material and that crater floors are rather constituted with water-ice rich materials, likely to probe the upper lithosphere of the moon (Solomonidou et al. 2018; Brossier et al. 2018; Werynski et al. 2019).

Mountainous features, in the form of ridges, blocks and chains, are found across Titan and indicate the presence, current or in the past, of internal stresses (Radebaugh et al. 2007; Mitri et al. 2010). They are rugged and heavily eroded, indicating exposure to erosional processes, similar to impact crater rims. Some mountains form long (>100 km) chains with undulatory planform morphologies, indicating the action of folding or faulting, most often in the E-W orientation (Radebaugh et al. 2007; Mitri et al. 2010). The general morphology and slopes from topography, where available, are consistent with an origin through compressional tectonism, which is rare on icy satellites but may be enabled by subsurface methane fluids, similar to contraction on Earth (Liu et al. 2016). Just how long, or how often, mountain-building processes occurred on Titan, and whether they are active today is unknown.

*3.4.2 Open questions*

Limited in coverage and in the highest achievable resolution, Cassini may have missed craters or mountains on Titan. The cumulative crater-size frequency distribution available today is likely to be rather incomplete, and the precise age of Titan's surface is still an open question. Also, craters and mountains provide invaluable windows into the crustal composition, still largely unknown. Additionally, their present-day morphology gives key information on the strength of surface erosion by winds and rain falls. The importance of unveiling the detailed compositional and morphological nature of impact craters and mountains, in order to characterize their state of degradation, lies therefore on the connection of the interior with the surface but also on the influence the atmosphere has on the surface.

Answering open questions regarding the impact craters on Titan remains a major goal for future missions to address:

- **What are the relative ages of all of Titan's geologic units?**
- **What is Titan's bedrock/crust composition?**
- **What are the erosion and degradation rates of craters and mountains? What do they reveal about Titan's past and present climate? What is the reason for the difference in the crater population of Xanadu Regio from other regions on Titan, and in particular for the paucity of craters in Titan's polar regions?**
- **How did the mountain belts of Titan form, for how long was tectonism active, and/or is it active today?**

*3.4.3 Proposed key instrumentation and mission concept to address those questions*

At the end of the Cassini mission, only ≈45% of the surface has been imaged by SAR at 300-1500 m resolution, and 20% of the surface has been seen by VIMS with a resolution better than 10km. Detailed geological investigations generally require at least under hectometer resolution, best would be decameter.

An **orbiter on Titan**, with high-resolution imaging capabilities (down to 10 meters) and overlapping use of instruments with infrared and microwave spectral capabilities would allow a systematic survey of impact and mountain features, providing constraints on the processes that have shaped the moon, the age of the surface, and the composition of the surface and subsurface, complementing the more detailed, but very local *in situ* exploration of the Selk crater and nearby mountain belts by Dragonfly. A **high-resolution 2-µm imager** would provide the best tradeoff between signal-to-noise ratio and atmospheric transparency. A near-polar orbit with a sufficiently long mission will guarantee global surface coverage. Again, the next mission to Titan should provide a **global "high-resolution" topography map** that is missing from Cassini, fundamental for Titan's geology study.

### 3.5 Interior-surface-atmosphere exchange processes

#### *3.5.1 Current knowledge*

Sub-surface ocean:

The Cassini-Huygens mission provided three independent pieces of evidence for a subsurface ocean at Titan (**Figure 10**). The first clue was provided by an unexpected measurement, made by the HASI-PWA instrument during the descent of the Huygens module into Titan's atmosphere, which revealed the existence of electrical disturbances (Simões et al. 2007). These electric signals were later interpreted as evidence of a Schumann-type resonance between two electrically conductive layers, the ionosphere at the top of the atmosphere and a deep layer, presumably composed of salt-containing water (Béghin et al. 2007), estimated at a depth of 50-80km below the surface (Béghin et al. 2010). The second piece of evidence was provided by the measurement of Titan's rotation, which is characterized by a tilt of 0.3° relative to the normal of its orbit (Stiles et al. 2010). Although low, this value is 3 times higher than what is expected for a solid interior and can be explained by the presence of a liquid layer decoupling the outer ice shell from the deep structure (Baland et al. 2011). The final proof was provided by measuring the temporal variations in Titan's gravitational potential, testifying of tidal fluctuations (Iess et al. 2012; Durante et al. 2018). The amplitude of the tidal fluctuations estimated from the dynamic Love number, $k_2$ ($k_2 \simeq 0.62 \pm 0.07$, Durante et al. 2019), can only be explained by the presence of a liquid layer under the ice layer (Iess et al. 2012; Mitri et al. 2014a). The high value of $k_2$ even suggests that it is a significantly denser layer than pure water, indicating a high salt content.

The Cassini-Huygens mission also provided key data on the long-wavelength topography (Zebker et al. 2009; Lorenz et al. 2013; Corlies et al. 2017) and low-degree gravity field of Titan, which put constraints on the 3-D structure of the ice shell (Choukroun and Sotin 2012; Hemingway et al. 2013; Lefevre et al. 2014; Mitri et al. 2014a; Kvorka et al. 2018). The long-wavelength topography is characterized by depression at the poles that could result rather from accumulation of heavy hydrocarbon clathrates (Choukroun and Sotin 2012) or thinning of the ice shell (Lefèvre et al. 2014; Kvorka et al. 2018). In the latter case, this would imply that the ice shell is in a conductive state in order to maintain significant ice shell variations (of the order of 5km) between the poles and the equatorial regions (Lefèvre et al. 2014) and that the ocean dynamics imposed a heterogeneous heat flux at the ice/ocean interface (Kvorka et al. 2018).

Cryovolcanism:

Since sunlight irreversibly removes methane at the top of Titan's atmosphere, the presence of 2-5% of methane for longer than ≈10-30 million years requires continued replenishment (Yung et al. 1984). The source of the atmospheric methane is one of the most outstanding mysteries on Titan. Outgassing by cryovolcanism has been proposed as a possible replenishment mechanism (e.g. Sotin et al. 2005; Tobie et al. 2006; Lopes et al. 2007) and plausible cryovolcanic landforms have been identified based on their morphology (Wall et al. 2009; Lopes et al. 2007, 2013) or evidence of change at the surface (e.g. Nelson et al. 2009; Solomonidou et al. 2016).

To date, the most convincing cryovolcanic region candidate includes Sotra Patera, a possible caldera consisting of a 1500-m-deep depression which is located next to Doom Mons, a mountain with a peak over 1000m high. Mohini Fluctus with its 100-m-thick lobate flows is visible on the flanks of the mountain and, farther north, another mountain over 1km high, Erebor Mons. Another cryovolcanic candidate is Tui Regio, a regional basin that includes interesting morphological features such as radar-dark and bright patches with sharp boundaries (e.g. Wall et al. 2009). Both Sotra Patera and Tui Regio have shown surface albedo fluctuations with time with pronounced trends for brightening and for darkening, respectively. The possibility also exists that explosive cryovolcanism created raised rim craters at Titan's north polar region (Mitri et al. 2019; Wood and Radebaugh 2020).

Interactions between ocean, rock core, and organic-rich crust:

Based on the gravity field measured by Cassini (Iess et al. 2012), Titan's hydrosphere is estimated to be about 500km thick (Castillo-Rogez and Lunine 2010; Fortes 2012) with the pressure at the rock/ice interface ranging between 0.7-1.0 GPa (Fortes 2012). Based on the water phase diagram, Titan's ocean is probably sandwiched between the outer ice crust and a deep mantle composed of high-pressure (HP) ice phases that separate it from the silicate core (**Figure 10**).

Another aspect concerns possible active exchange processes between the organic-rich crust and the ocean. Interaction of the organic-rich crust with the underlying water ocean due to large impacts (Lunine et al. 2010; Zahnle et al. 2015) or Rayleigh-Taylor convective instabilities have been proposed, even though there are still no observational constraints of potential recycling.

Apart from the surficial indications, it is suggested that liquid pockets with methane clathrates and with a high ammonia mass concentration in a water solution can dissociate in the ice shell and eventually exsolve on the surface and in the atmosphere (Tobie et al. 2006; Mitri et al. 2008). Hence, cryovolcanism can act as the dynamic force that deforms tectonic features and brings methane into the atmosphere. However, the source of heat that causes cryovolcanism remains unclear. A number of heat sources have been proposed for Titan, such as radiogenic heating, and tidal dissipation, since Titan is subject to solid body tides exerted by Saturn on the time-scale of its orbital period (e.g. Tobie et al. 2005; Iess et al. 2012). The occurrence of dynamical processes such as cryovolcanism largely depend on the tide-induced internal redistribution of mass that results in variations of surface gravity, tilt, stress, and strain. Modeling of Titan's tidal potential and correlation with Titan's surface features have suggested that the concentration of the major cryovolcanic candidates around the equator and their young geological age could connect dynamic movements to surficial stresses related to Titan's tidal environment (Sohl et al. 2014).

Titan's atmosphere formation and evolution:

A preliminary requirement for assessment of the astrobiological potential of Titan is to constrain the origin(s) of volatile compounds and to determine how their inventory evolved since satellite accretion. The present-day composition of Titan's atmosphere, as revealed by Cassini–Huygens, results from a combination of complex processes including internal outgassing, photochemistry, escape and surface interactions. The detection of a significant amount of $^{40}Ar$ (the decay product of $^{40}K$) by Cassini–Huygens (Niemann et al. 2005, 2010; Waite et al. 2005) indicated that a few per cent of the initial inventory was outgassed from the interior.

The isotopic ratios in different gas compounds observed on Titan constitute crucial constraints to assess their origin and evolution. Cassini-Huygens and ground-based measurements provided isotopic ratios of H, C, N, and O in $N_2$, CO, $CH_4$, HCN, and $C_2$ hydrocarbons at various altitudes in Titan's atmosphere (e.g. Mandt et al. 2012; Nixon et al. 2012). The measured $^{15}N/^{14}N$ ratio is enigmatic because it is about 60% higher than the terrestrial value (Niemann et al. 2010), suggesting an abnormally high fractionation. In contrast, $^{13}C/^{12}C$ in methane implies little to no fractionation, suggesting that methane has been present in the atmosphere for less than a billion years (Mandt et al. 2012). In the absence of a proper initial reference value, however, it is impossible to retrieve information on fractionation processes with confidence. Precise isotopic ratios in the photochemical by-products of $CH_4$ and $N_2$ on Titan are also lacking (see Dobrijevic et al. 2018).

The salt enrichment assessed in the ocean (Mitri et al. 2014a; Tobie et al. 2014) as well as the $^{40}Ar$ atmospheric abundance (Niemann et al. 2010) suggests an efficient leaching process and prolonged water-rock interactions which may have affected the volatile inventory and possibly may explain the main atmospheric gas compound (Tobie et al. 2012; Glein 2015).

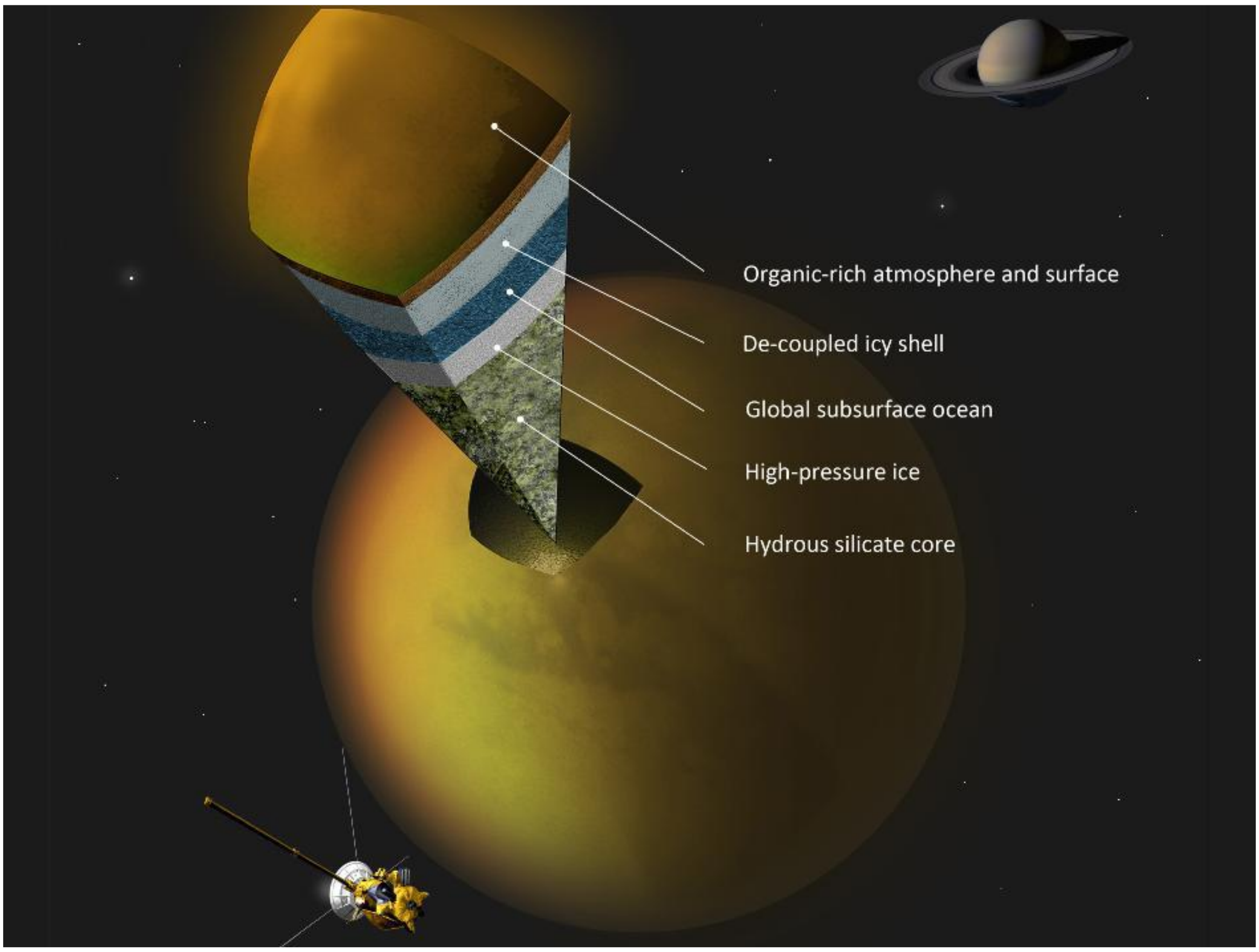

*Figure 10: Titan's internal structure (*Credits: NASA/JPL-Caltech/Space Science Institute).

### *3.5.2 Open questions*

Despite the unprecedented advances in our knowledge of Titan's internal structure accomplished thanks to the Cassini-Huygens mission, the existence of a subsurface global ocean and the possible exchange between Titan's interior, surface, and atmosphere, numerous fundamental questions are still unanswered, due to a lack of specific measurements or to still too uncertain measurements.

For instance, even if the current uncertainties on the estimation of $k_2$ are relatively good, it is not possible to conclude with certainty about the density and depth of the ocean.

In the same manner, although likely nowadays, the presence and thickness of the HP ice mantle are uncertain throughout Titan's evolution as they depend on two highly unknown quantities - the thickness of the ice crust and the salinity of Titan's ocean (Kalousova and Sotin 2020). Depending on the mantle thickness, several HP phases may be present – ice VI, ice V, and even ice III if the mantle is more than 250 km thick (Kalousova and Sotin 2020). Numerical simulations indicate that the dynamics of this HP ice mantle that governs the heat extraction from Titan's silicate core and potential chemical exchange from the core to the ocean through the upward advection of aqueous fluids strongly depend on the HP mantle thickness, ice viscosity, and ocean composition (Choblet et al. 2017; Kalousova et al. 2018), which still remain poorly constrained based on Cassini-Huygens data.

The fundamental question of whether cryovolcanism takes place (or has taken place in a recent past) on Titan is still debated (e.g. Moore and Pappalardo 2011) and the chemical exchanges with the surface and the interior, as well as the initial composition, still remain unconstrained (e.g. Tobie et al. 2014).

Our comprehension of the mechanism governing the formation and evolution of Titan’s atmosphere rely on precise determination of isotopic ratios in N, H, C, and O-bearing species in Titan's atmosphere and in surface materials (organics, hydrates, and ices), which are lacking. They will permit a better determination of the initial reference ratio and a quantification of the fractionation process due to atmospheric escape and photochemistry.

Finally, after Cassini, the chemical exchanges associated with water-rock interactions, conditioning the composition of the internal ocean, are only hypothesized. They may be quantified by accurately measuring the ratio between radiogenic and non-radiogenic isotopes in noble gases (Ar, Ne, Kr, Xe) in Titan's atmosphere (Tobie et al. 2012).

To summarize, the remaining open questions are:

- **What are the depth to, volume, and composition of the subsurface liquid water ocean?**
- **Is Titan currently, or has it been in the past, cryovolcanically active?**
- **Are there chemical interactions between the ocean, the rock core, and the organic-rich crust?**
- **How did Titan's atmosphere form and evolve with time in connection with the interior?**

*3.5.3 Proposed key instrumentation and mission concept to address those questions*

A combination of geophysical measurements from the orbit (**radio experiment, radar altimeter and sounder, radar imager, magnetometer, and plasma package**) and from the ground (**seismometer**, **radio transponder** and **electric sensors, and magnetometer**) is required to constrain the hydrosphere structure, and hence provide essential constraints on the ocean composition and the thickness of both the outer ice shell and the high-pressure ice mantle (if any). This is a preliminary requirement for assessment of the astrobiological potential of its internal ocean. **Surface sampling** of erupting materials could reveal the salt composition of the ocean.

Measurement by a **high-precision mass spectrometer** of the ratio between radiogenic and non-radiogenic isotopes in noble gases (Ar, Ne, Kr, Xe) in Titan's atmosphere will also put fundamental constraints on whether and how water-rock interactions have occurred in Titan's interior. Comparison between isotopic ratios in N, H, C, and O-bearing species in the atmosphere (gas and aerosols) and in collected surface materials, at different locations, will provide key information on the volatile origin, if they were chemically reprocessed in the interior and/or at the surface.

Future missions could have on board a **high-resolution microwave radiometer** to look for thermal anomalies (hot spots) revealing possible still active cryovolcanism on Titan, which cannot be observed in optical and infrared domains due to atmospheric opacity. Microwaves also offer the prospect of sensing the shallow subsurface and thus may detect warmth from old lava flows, i.e. lava flows which have cooled at the surface and thus have no more infrared emission signature but are still tens of K above ambient at depth. An **infrared spectrometer of higher spectral resolution**, especially at 1.59µm, 2µm, 2.7µm and 5µm, and overlapping capabilities with a **radar instrument** and **high-spatial resolution infrared camera** would help with the identification of specific constituents (such as $NH_3$ or local enrichment in $CO_2$) and their spatial distribution that could only have internal origin, and thus function as an additional “smoking gun” for cryovolcanism on Titan. Detailed mapping of geomorphological features, composition, topography, and subsurface sounding in regions of interest may also reveal areas where recycling processes have occurred.

## 4. Science Goal C: Titan's habitability

### *4.1 Current knowledge*

Habitable environments are defined as favoring the emergence and the development of life (Lammer et al. 2009). Habitability is based on the presence of a stable substrate, available energy, organic chemistry, and the potential for holding a liquid solvent. This definition identifies Titan as one of the celestial bodies in the Solar System with the highest potential for habitability. Titan harbors a complex organic chemistry producing a plethora of molecules and organic haze particles, and several sources of energy are available: solar radiation, solid body tides exerted by Saturn (Sohl et al. 2014), radiogenic energy production in the body core regions (Fortes 2012), and possible exothermic chemical reactions may be at work (Schulze-Makuch and Grinspoon 2005). Strikingly, two kinds of solvents are simultaneously present on Titan: (1) **a dense, likely salt-rich, subsurface ocean with an unconstrained fraction of ammonia** (Tobie et al. 2006; Iess et al. 2012; Béghin 2015), (2) **a mixture of simple hydrocarbons in liquid state**, forming a collection of seas and lakes in the polar regions (Stofan et al. 2007). These circumstances offer the possibility of the existence of **two distinct possible biospheres** between which fluxes of matter could be established, via geological processes like cryovolcanism. These chemical transfers are supported by the detection of $^{40}Ar$, the decay product of $^{40}K$ initially contained in rocks from the core, in the atmosphere (Niemann et al. 2010). Our knowledge about Titan's habitability, even if several interesting conceptual works have been published (McKay and Smith 2005; McKay 2016), remains very poor and essentially speculative. An on-site investigation is therefore invaluable for improving our views.

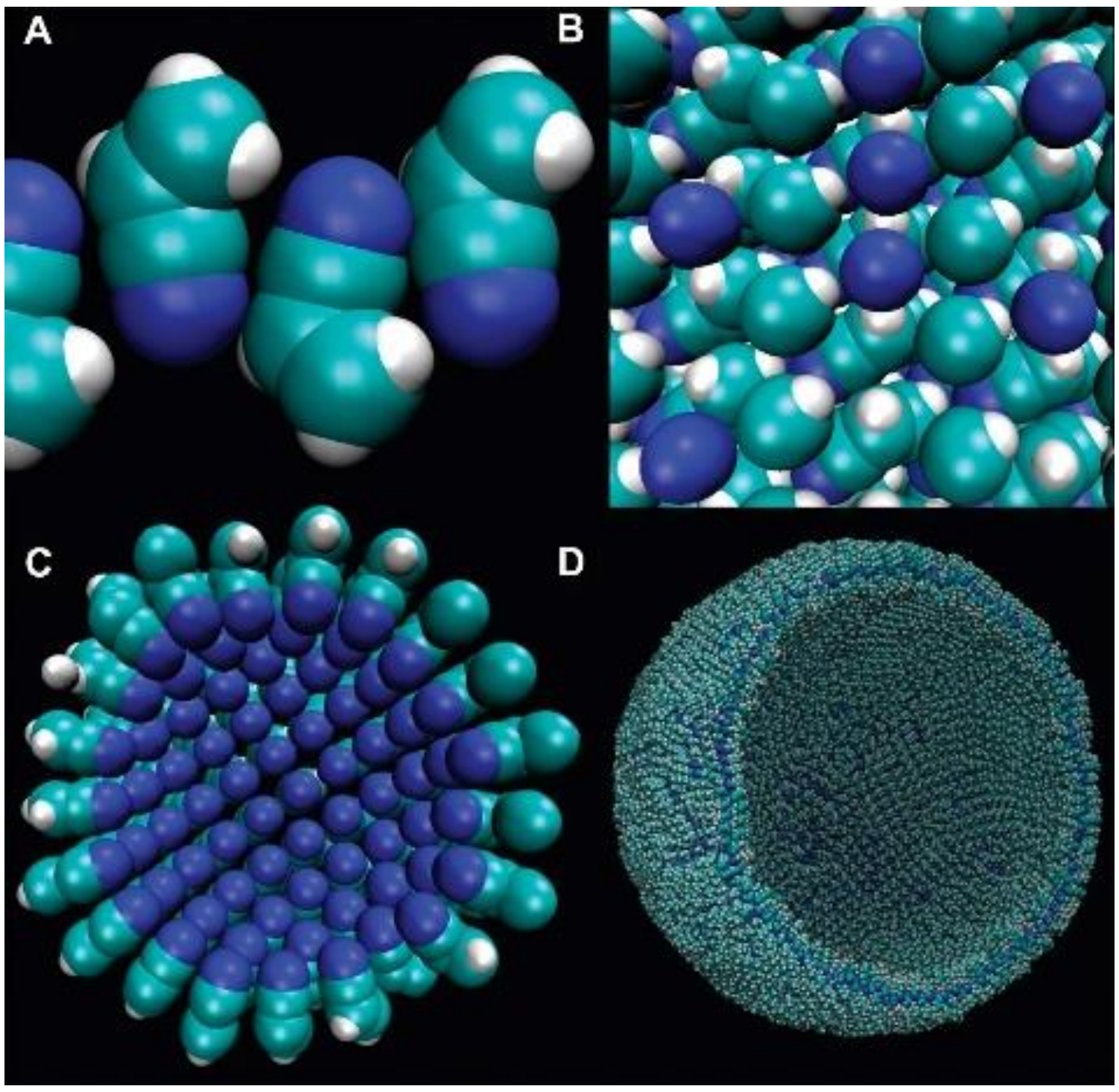

Figure 11: Proposed structure of Titan "azotosome": a membrane that may be stable in liquid methane composed of acrylonitrile sub-units. Enclosed membranes are considered a vital prerequisite for life (Stevenson et al., 2015a,b). Image: Cornell U./Science Advances.

*4.2 Open questions*

Clearly, a potential "**aqueous Titan biotope**", hidden below the icy crust, is too deep to be directly explored. However, as already presented in **Section 3.5**, a series of geophysical measurements and chemical analysis of freshly erupted surface samples may provide crucial information on the ocean composition, on possible water-rock-organic interactions and internal heat sources. Besides, liquid water can be brought to the surface by cryovolcanism (Lopes et al. 2007) or cratering events (Artemieva and Lunine 2003). Once deposited on Titan's surface and in contact with liquid water, complex organic content produced by the atmosphere may lead to the production of biologically important species such as amino acids and purines (Poch et al. 2012). Long-term chemical evolution is impossible to mimic experimentally in the laboratory. It is, therefore, crucial to be able to perform a detailed *in situ* chemical analysis of the **surface zones where cryovolcanism and impact ejecta (or melt** **sheets) are or have been present**, by using **direct sampling by drilling the near subsurface and/or performing spectroscopic observations**. Indeed, finding water on Titan's surface or near-surface at shallow depths is important to understand oxidation processes of surface organics that could lead to biologically important molecules. Spectroscopic observations augmented by drilling through shallow depths of ≈50 cm could provide immediate answers to these chemical reaction pathways.

Lakes and seas of hydrocarbons are also environments that could host a potential "**liquid hydrocarbon biosphere**". The exact chemical composition of these systems is also debated (Cordier et al. 2009, 2013; Luspay-Kuti et al. 2012; Tan et al. 2013), but it is generally well accepted that the liquid should be a mixture of three main species: methane, ethane and nitrogen, in variable proportions. A direct sampling is required to get a more solid picture of this composition (see **Section 3.3**). The role of these cryogenic solvents (the composition could be variable in time and space) is twofold: (1) liquids may interact with materials into which they are in contact, (2) the bulk volume may harbor a multitude of chemical processes. Probably the first question arising is the possible interaction between these cryogenic liquids and complex aerosols falling from the atmosphere. The transparency at the radar wavelength, of most of the lakes, indicates probably a low aerosol content (Mastrogiuseppe et al. 2019). However, the exact aerosol content of these lakes as well as the amount that could have sedimented at the bottom of the lakes still remain unconstrained. The presence of a floating film has been already proposed (Cordier and Carrasco 2019), the existence of such a deposit could be easily detected by a **Titan lake lander or a drone with the capability to land and float on liquids (hydrodrone)**. The determination of its nature (monomolecular layer? thicker deposit made of aerosols?) is also important: the presence of a monomolecular layer could be the sign of the existence of a kind of "surfactant" which may have biological implications. Indeed, these classes of molecules form micellas which could be the first stage of the formation of "vesicles", in which a specific "proto-biochemistry" could appear. In a world with very little oxygen like Titan, analogs to terrestrial liposomes, entities called "azotosomes" (**Figure 11**) have been already theoretically studied (Stevenson et al. 2015a,b) while their main compound, $C_2H_3CN$, has been recently detected in Titan's atmosphere (Palmer et al. 2017). This also reinforces the need for accurate chemical analysis of the liquids in Titan's seas, beyond the "simple" determination of $N_2$+$CH_4$+$C_2H_6$ mixing ratios.

Naturally, the ocean-atmosphere interface could reveal a large physico-chemical processes diversity. Recently, as an interpretation of the "Magic Islands" events (Hofgartner et al. 2014, 2016), bubbles streams coming from Ligeia Mare depths have been proposed (Malaska et al., 2017; Cordier et al. 2017; Cordier and Liger-Belair 2018). This scenario could be confirmed by direct investigations, while it is important for exobiological activity since it implies the stability of the solvent and the sea/lake global mixing of material. **The exploration of these seas and lakes requires a multi-instruments approach going from global imaging, to more specific *in situ* chemical measurements** or indirect ones performed by specific instruments like those based on sound speed measurements (Cordier 2016).

**In the fringes of Titan's lakes, possible evaporite deposits** may exist, which is supported by infrared observations combined with radar imaging (Barnes et al., 2011), and also by numerical models (Cordier et al. 2013, 2016). According to the latter, due to its solubility properties and abundance, acetylene could be the dominant component of these evaporitic layers. From an astrobiological perspective, a massive presence of acetylene is of prime interest. Indeed, at the surface of Titan the total solar radiation flux is only ≈0.1% of its terrestrial counterpart (Tomasko et al. 2005, 2008b). For this reason, Titan is not favoured compared to the Earth. The quantification of this potential chemical energy source would require the measurement of the abundances of acetylene (Strobel 2010) at the surface of Titan. If acetylene has been identified in the atmosphere (Coustenis et al. 2007); its presence, in solid state, in drybeds, or lake beaches is not clear, even if it has been possibly detected in the equatorial dune material (Singh et al., 2016). **Drilling an evaporite field could bring numerous crucial constraints concerning the replenishment/evaporation cycle of Titan's lakes, together with interesting clues about potential chemical feedstocks and their transport across the Titan surface.**

On Earth, a favored chiral structure is a characteristic property of biogenic molecules. Then, detection of an imbalance in the abundance of different handed chiral molecules may be used as an indicator of some biological activity on Titan. In a terrestrial context, it has been emphasized that living beings use chiral, stereo-chemically pure macromolecules (Plaxco and Allen 2002). These molecules show a noticeable circular dichroism in the terahertz domain, which could be used as a general biosignature (Xu et al. 2003). To **determine chiral properties of samples collected** on the surface of Titan **requires the development of a dedicated instrument**, such as COSAC on Philae, MOMA on ExoMars, and DraMS-GC on board Dragonfly (here complementing the measurements of Dragonfly by sampling and analysing material from the polar seas/lakes environment).

The numerous remaining open questions regarding the potential habitability of Titan are summarized here:

- **What is the nature and quantity of material exchange between the subsurface ocean and the surface? In the past, did a form of life develop in the water pond, formed by cryovolcanism or bolide impacts?**
- **How is the organic material falling from the atmosphere physically/chemically processed at the surface? Does some catalytic path exist for the hydrogenization of acetylene or other reactions? How prevalent is water ice on Titan's surface? What is the depth of organic deposit on the ice (if measurable by drilling and/or radar)?**
- **Does a layer of surfactant (or even thicker deposit) cover the surface of some lakes/maria?**
- **What is the nature of dissolved species in hydrocarbon lakes? Does this liquid environment harbor a chemical reactions network?**
- **Are the molecules present in lakes and evaporites deposits optically active? Can a kind of homochirality be exhibited?**

### *4.3 Proposed key instrumentation and mission concept to address those questions*

A very high-resolution mass spectrometer is needed, like the already mentioned **CosmOrbitrap**, that should be used for low atmosphere composition measurement. An instrument of the same class is also needed for the analysis of liquid phases and solid surfaces (evaporitic terrains and crater soils), complemented by specific samplers for both phases. These **samplers should be a drill and an instrumented diving probe**, linked to a main "sea lander" (lake lander or amphibious drone) by a technical cable supplying power and commands, and collecting measurement signals. Concerning chirality determinations, a **chiral gas chromatograph** (Patil et al. 2018) will be suitable.

## 5. General mission concepts

### 5.1 Previous mission concepts for post-Cassini-Huygens exploration of Titan

Post-Cassini exploration of the Saturnian system with a focus on Titan has been considered for quite some time, almost since the earliest years of the Cassini-Huygens mission. The Titan explorer (Leary et al. 2008) and the Titan and Enceladus Mission (TandEM, Coustenis et al. 2009) concepts had been selected respectively by NASA and ESA for studies before they were merged into the joint large (Flagship) Titan and Saturn System Mission (TSSM) concept, which was extensively studied in 2008 (Reh et al. 2008; K. Reh, C. Erd, D. Matson, A. Coustenis, J. Lunine and J.-P. Lebreton, ESA, TSSM NASA/ESA Joint Summary Report, 2009[2]). TSSM aimed at an in-depth long-term exploration of Titan's atmospheric and surface environment with a dedicated orbiter, and *in situ* measurements in one of Titan's lakes (with a lake lander) and in the atmosphere with a montgolfière (hot air balloon).

TSSM was ranked second in the final decision by the agencies and was not considered for further study. It is still, however, one of the most ambitious mission concepts dedicated to Titan exploration to date, and has inspired several other proposed concepts, aborted or not selected, for smaller size programs and different payloads: **Titan Aerial Explorer** (TAE), a pressurised balloon (Hall et al. 2011); **Aerial Vehicle for *in situ* and Airborne Titan Reconnaissance** (AVIATR), an ASRG (Advanced Stirling Radioisotope Generator) powered airplane (Barnes et al. 2012); **Titan Mare Explorer** (TiME), a lake lander (Stofan et al. 2010); **Titan Lake Probe**, which included a submarine concept (Waite et al. 2010);

[2] https://sci.esa.int/web/tandem-tssm/-/44033-tssm-nasa-esa-joint-summary-report

**Journey to Enceladus and Titan** (JET), a single Saturn orbiter that would explore the plume of Enceladus and the atmosphere and surface of Titan (Sotin et al. 2011); a **seismic network** had been considered as part of the geophysical payload of such missions (Lorenz et al. 2009); **mission concepts with two elements: a Saturn-Titan orbiter and a Titan Balloon** (Tobie et al. 2014) and **a Saturn-Titan orbiter and a lake probe** (Mitri et al. 2014b); a **Saturn-Titan orbiter** (OCEANUS) (Sotin et al., 2017).

Among the most recent proposals, Dragonfly (Lorenz et al. 2017), an extraordinary and inspiring mission concept, involving for the first time the use of an autonomous rotorcraft to explore *in situ* the surface and low atmosphere of Titan, has been selected by NASA in June 2019 as its 4$^{th}$ New Frontiers mission. Dragonfly is scheduled for launch in 2027, arriving at Titan by 2034. The spacecraft will touch down in dune fields in the equatorial regions, near the Selk crater. From there Dragonfly will fly from location to location covering a distance of up to ≈175 kilometers in its 3.3-year nominal mission. Despite its unique ability to fly, Dragonfly would spend most of its time on Titan's surface making science measurements (sampling surface material and ingesting into a mass spectrometer to identify the chemical components and processes producing biologically relevant compounds; measuring bulk elemental surface composition with a neutron-activated gamma-ray spectrometer; monitoring atmospheric and surface conditions, including diurnal and spatial variations, with meteorology and geophysical sensors; performing imaging to characterize geologic features; performing seismic studies to detect subsurface activity and structure). In-flight measurements are also planned (contributing to atmospheric profiles below 4km in altitude, providing aerial images of surface geology, and giving context for surface measurements and scouting of sites of interest). Unable to use solar power under Titan's hazy atmosphere, Dragonfly is designed to use an MMRTG (Multi-Mission Radioisotope Thermal Generator). Flight, data transmission, and most science operations will be planned to occur during Titan's daytime hours (≈eight Earth days), giving the rotorcraft plenty of time during the Titan night to recharge its battery. Dragonfly is a great step forward in Solar System exploration history, technologically and scientifically speaking, pioneering the use of an extremely mobile and comprehensively instrumented *in situ* probe to investigate the atmosphere-surface-interior interactions of Titan. NASA and Dragonfly are paving the way to an ambitious, international program to explore the extreme complexity of Titan, hopefully in synergy and partnership with other agencies including ESA.

In May 2019, a mission concept (TOPS: Titan Orbiter/Polar Surveyor[3]) was submitted to NASA for consideration as a major flagship study for the forthcoming Decadal Survey for Planetary Sciences, expected in 2021. This mission concept proposes an instrumented orbiter, and also 1-2 probes to land on Titan's largest lake(s): Kraken and/or Ligeia. If selected, the TOPS mission would allow for significant synergy with the Dragonfly mission and involvement by international partners including ESA, with potentially contributions of instruments, subsystems, or launch, subject to the appropriate bilateral agreements being negotiated. This would allow continuation of the scientific collaborations that made Cassini-Huygens so successful, and widen the participation of the international scientific community.

### 5.2 An ESA L-class mission concept for the exploration of Titan

In order to fulfil the totality of the key science goals presented in the present article, we estimate that, in response to the ESA Voyage 2050 call for proposals, an ESA L-class mission involving a Titan orbiter and at least an *in situ* element that has sufficient mobility to probe the atmosphere **and** the solid and possibly liquid surface (thus excluding balloon) is required. Inspired by the experience of TSSM, such a mission, with international collaboration to support the overall architecture and cost, will perfectly complement, and surpass, the exploration of Titan undertaken in the 2000s by the NASA-ESA-ASI Cassini-Huygens mission and reactivated by the selection of the NASA Dragonfly mission concept and the TOPS proposal.

Titan's equinox would be the ideal season for observing tropical storms and their consequences for fluvial and aeolian features. It is also the best period to observe strong changes in the global atmospheric dynamics and its impact on the distribution of photochemical compounds, as well as on the thermal field. Equinoxes on Titan during the ESA Voyage 2050 period will be on 22 January 2039

[3] https://www.lpi.usra.edu/opag/meetings/feb2020/presentations/Nixon.pdf

(northern Spring equinox) and on 10 October 2054 (northern Autumn equinox). Having a mission with an orbiter planned to monitor the seasonal transition over the 2039 northern Spring equinox would have the extraordinary opportunity to potentially overlap with the Dragonfly mission, complementing its scientific targets and even possibly acting as a transmission relay. We therefore advocate for a launch window in the early phase of the ESA Voyage 2050 cycle or earlier. Considering an estimation of the cruise from Earth to Saturn to 7-8 years, a launch as early as 2031-2032 would be required. In the case of a partnership of ESA with NASA regarding the Dragonfly mission, our arrival at Titan should be as early as 2034. The clock is ticking! Note that any arrival outside those dates would follow on the results of Dragonfly and would still have, with the use of an orbiter and an *in situ* element, an outstanding scientific value, still answering fundamental open questions that remain about Titan's system that cannot be answered from terrestrial ground-based or space-borne facilities.

We propose that, upon arrival, the **orbiter** will be captured around Titan in an elliptical orbit followed by a few months of aero-braking. This aero-braking phase will enable the exploration of a poorly known, but chemically critical, part of the atmosphere (700-800km), with *in situ* atmospheric sampling at altitudes much lower than possible with Cassini. Following the aero-braking phase, the orbiter will be placed into a near-polar elliptical orbit of lower eccentricity for the orbital science phase of a nominal duration of at least 4 years. Orbital periapse will be located in the thermosphere to continue performing *in situ* mass spectra measurements for each orbit. It will also allow imagery and spectroscopy of the surface and atmosphere at all latitudes with high spatial and temporal resolution, and repeat coverage over time.

While an orbiter would be of high value to provide global coverage, we have demonstrated above and are deeply convinced that the addition of at least one ***in situ* probe** is critical in terms of scientific complementarity with the orbiter for any future ambitious mission to Titan (following the legacy of the extraordinary Cassini-Huygens mission). In addition to the orbiter, which can serve as a transmission relay, we thus propose a mission scenario with ***in situ* element(s) to explore the polar regions of Titan**, complementary to the Dragonfly mission: a lake lander, a drone and/or an air fleet of mini-drones.

Titan's low gravity and dense atmosphere make it an ideal candidate for drone-based missions. The *in situ* probe(s) will be able to also perform atmospheric measurements directly inside the polar vortex and image the surface during the descent phase.

We list below the possible *in situ* probe scenarios:

- A **Titan lake lander** was previously proposed in two mission proposals and one mission concept (Reh et al. 2008; Stofan et al. 2010; Waite et al. 2010). Both mission candidates benefited from extended studies. Less mobile and more scientifically focused than a drone, this type of mission element, and corresponding mission architecture, is less complex and risky. A Titan lake lander can directly be built on ESA's heritage established with the Huygens probe and does not require critical new technological developments. The lander would preferentially float and passively drift for days at a lake's surface (with long-lived batteries), possibly including the additional capability of plunging. Such a probe will directly sample and observe the lake's liquid properties (temperature, viscosity, permittivity, composition, wave/current activity, and tidal deformation), as well as its geological context (shorelines and surroundings' composition and morphology, depth) and local meteorological conditions, which cannot be done from orbit, or only with extreme difficulty, larger uncertainties, and lower resolution.

- A **Titan heavy drone** is a much more capable alternative to the immobility (or very low mobility) of a lake lander and **is our preference**. Also proposed by the AVIATR consortium, the idea to use an autonomous flying drone to explore Titan's low atmosphere and surface is an original concept and thus requires the resolution of numerous challenges and the development of numerous innovative technologies. The feasibility study of the AVIATR unmanned aircraft and the selection by NASA of Dragonfly (a Mars' rover-sized octocopter) promise near-future operational drones for planetary exploration to all agencies and to the entire scientific community. In perfect complement to the timing, area, and topics of exploration of Dragonfly, a flying drone in the North Polar regions, arriving before northern Spring equinox, would provide an extraordinary tool for exploring the complex geology and meteorology of the lake's district in a context of changing seasons. Equipped with MMRTGs, the "lake" drone would be able to fly from one lake to another, and land close to their shorelines. It could thus directly sample and measure regional variations in composition and geology, in local wind, humidity, pressure, and precipitation, and how the climatology interacts with the liquid bodies, with unprecedented precision. If we add the ability to fly up to 10km altitude and also land and float a few minutes on lakes (and sample their liquids), a hydrodrone, would have the ideal and most comprehensive *in situ* element concept to explore the geology and lower atmosphere of Titan's polar region.

- A **Titan air fleet of mini-drones** would be an interesting and complementary alternative concept to a single drone, simultaneously guaranteeing a larger range of action and the concomitant study of multiple targets, without risking the single-point failure mode of a large drone. Again, we propose this concept to explore the polar regions of Titan in order to complement the scientific goals of the Dragonfly mission. While significantly enhancing the scientific return of the mission for a reasonable additional cost, mini-drones would have to face significant thermal challenges under Titan's conditions, and crude limitations of their range of action and data volume transfer (Lorenz et al., 2008c). A way to mitigate those restrictions would be to escort the mini-drone fleet with a lander (the proposed lake lander or a companion station based on the solid surface) or a mobile mothership (the proposed "heavy" drone) from which the mini-drones could take off and dock back. Such a (mobile or immobile) platform could be deployed at the North Polar region and could serve as a communication relay with the mini-drones and the orbiter, and as a possible recharging and heating station for a greater longevity and range. It could also serve as an analysis base where mini-drones could deposit samples. The mini-drones (e.g. mono-copter cubes of a few inches in size), that could be inspired from the "Mars Helicopter Scout" (Ingenuity), a mini-copter that flew to Mars with the Mars 2020 rover (Perseverance), would naturally be less capable than a large drone, but could nevertheless fulfil a wealth of observations and measurements by including a well-focused miniaturized payload (cameras, meteorological and electric environment package, simple solid/liquid sampling device, etc). The mini-drones could also interact and organize in formation flying if they were close enough. For instance, a fleet of four mini-drones could fly in different types of formation: 4x1 cross-track formation in order to enhance lateral coverage and mapping, 1x4 along-track formation (like the A-train constellation) in order to monitor at high time frequency the local meteorology and also provide stereo-imaging and give access to local meso- to micro-topography (complementary to macro-topography achievable from orbit), and 2x2 formation for a trade-off between mapping and stereo-imaging (**Figure 12**). During the course of the northern mini-drones' mission, an additional short-lived mini-drone could be released by the orbiter to explore the large South Pole lake, Ontario Lacus, providing a unique opportunity to investigate *in situ* two poles at the same time.

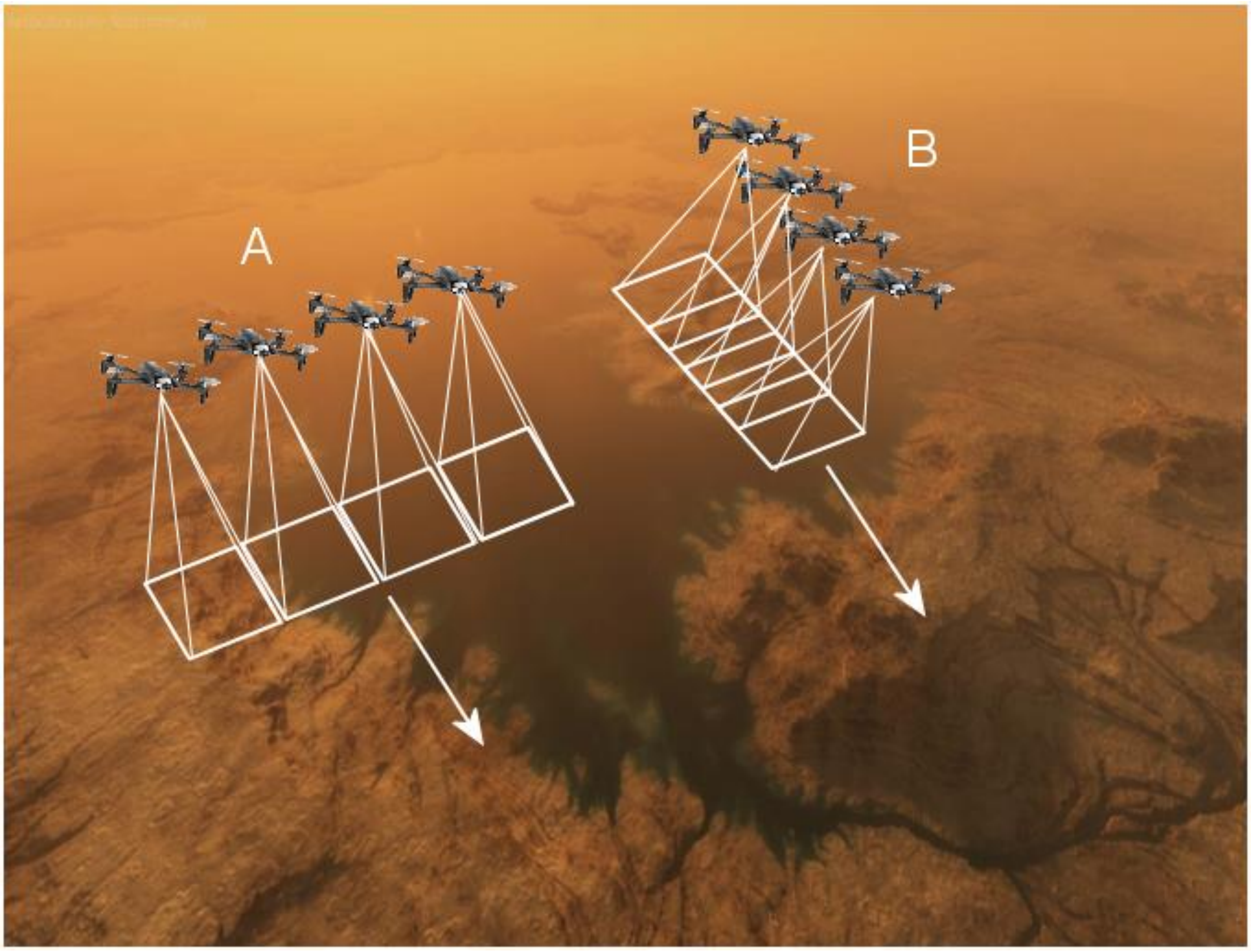

Figure 12: Bird's-eye view of the hydrocarbon sea Ligeia Mare, North Pole of Titan. The topography is extracted from Cassini RADAR SAR images and textured using the same set of images. The view has been realistically coloured and illuminated (Credits: Université de Paris/IPGP/CNRS/A. Lucas). Illustrated are two possible configurations for the flying formation of four mini-drones: (A) 4x1 cross-track and (B) 1x4 along-track enabling stereo-imaging at very high resolution.

We could add to those *in situ* concepts two very revolutionary, but preliminary technologies whose study has been supported by the NASA Innovative Advanced Concepts (NIAC) program and which are quite relevant for the exploration of icy moons in general and Titan in particular. These are the SPARROW (Steam Propelled Autonomous Retrieval Robot for Ocean Worlds) project, an autonomous mini-robot powered by steam that can hop over some of the most hazardous terrains known (and unknown) of the icy moons[4]; and the ShapeShifter project, a versatile self-assembling robot made of smaller robots that can separate and re-assemble to change shape and function, offering the possibility to roll, fly, float, and swim[5].

All those *in situ* probe concepts would be highly complementary (in terms of targets, measurements and/or achievable resolution) to the orbiter, but also to all available observatories and space missions that will operate at (or close to) the time of the ESA Voyage 2050 program (Dragonfly, JWST, ALMA, large Earth-based telescopes with Adaptive Optics…).

### 5.2.1 Strawman instrument payload

**Tables 1 and 2** present a tentative payload that would address the required measurements for the science goals A, B, and C. The proposed instruments will benefit from the heritage of successful missions such as Cassini-Huygens, Rosetta, Venus Express and Mars Express, as well as new missions currently under development (such as JUICE, ExoMars, …).

We propose in particular that one of the new *in situ* key instruments to be carried on the orbiter and the "heavy" *in situ* element(s) will be a CosmOrbitrap (Briois et al., 2016), which will acquire mass spectra with unequalled mass resolution. This instrument has a mass resolution 100 times higher than the mass spectrometer that will be on board Dragonfly (a Curiosity/SAM or ExoMars/MOMA type spectrometer). **Figure 13-left** shows the interest of high mass resolution to determine without ambiguity the composition of a gas, solid, or liquid. The CosmOrbitrap is currently TRL 5[6]. It has an expected mass of 8kg and needs 50W of power, and therefore could be included in a "heavy" drone/lander payload.

[4]https://www.jpl.nasa.gov/news/news.php?feature=7686&utm_source=iContact&utm_medium=email&utm_campaign=nasajpl&utm_content=daily-20200624.3
[5] https://www.jpl.nasa.gov/news/news.php?feature=7505

Significant new insights into Titan's atmosphere and surface composition will also come from the use of a near-infrared hyperspectral imager with a major increase in spectral range (especially above 5µm where numerous organics have diagnostic absorption bands (Clark et al., 2010)), and spectral and spatial resolutions. This is exemplified in **Figure 13-right**, where the ethane 2-µm absorption band diagnostic structure becomes apparent at spectral resolutions 10x to 30x times the resolution of Cassini/VIMS.

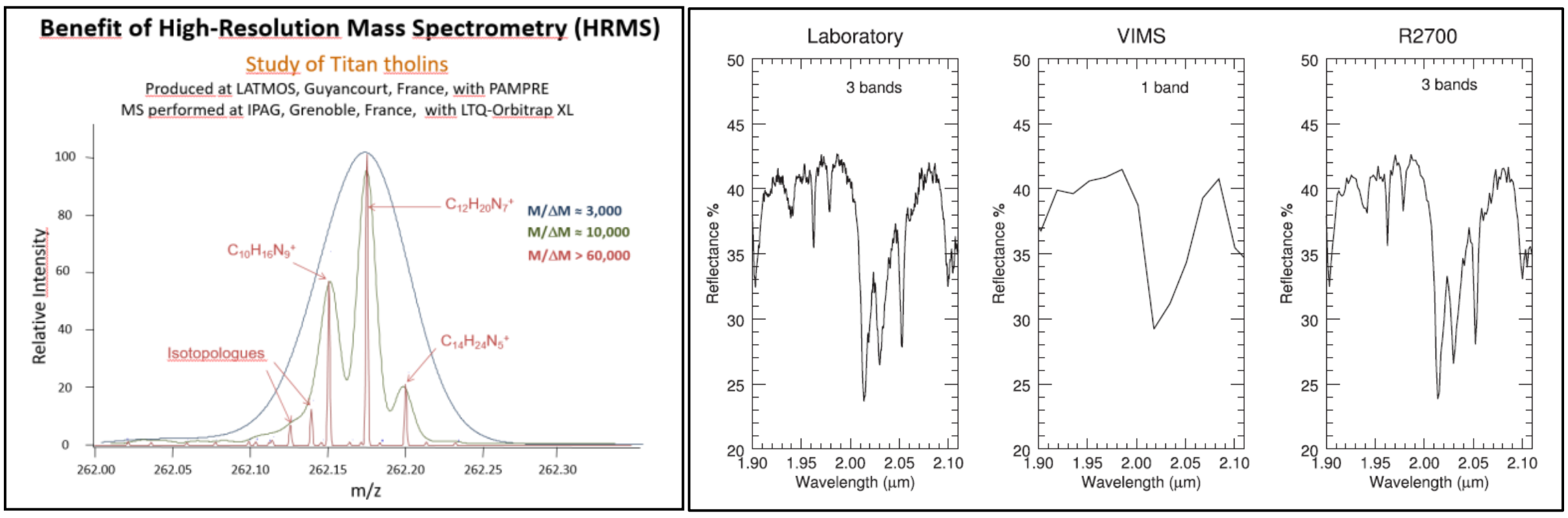

Figure 13: (Left) Mass spectrum of aerosol analogues (tholins) acquired at different mass resolution. CosmOrbitrap will have a resolution M/ΔM > 60,000 (the Cassini/INMS mass spectrometer had a resolution of 500). (Right) Comparison of a laboratory infrared spectrum of liquid ethane at 2.0µm, with the same absorption viewed at Cassini/VIMS spectral resolution (≈100-300) and at spectral resolution 2700. The diagnostic triple band at 2 µm shows up only at high spectral resolution (>1000). Laboratory spectra from the Arkansas Center for Space and Planetary Sciences.

[6] Technology Readiness Level (see https://sci.esa.int/web/sci-ft/-/50124-technology-readiness-level)

*Table 1: Tentative instrument payload to address the three mission goals A, B, and C.*

| ***Titan Orbiter*** | ***Titan probe (lander, drone(s))*** ***Possible payload for mini-drones is indicated in Tblue*** |
|---|---|
| 1. High spatial resolution imager (2μm, 2.7μm, 5-6μm) and high spatial and spectral resolution (R>1000) near-IR Spectrometer (0.85-6μm) **[A,B,C]** | 1. Visual Imaging System (two narrow angle stereo cameras and one wide angle camera) **[A,B,C]** |
| 2. Radar active and passive imager **[B,C]** | 2. Imaging Spectrometer (1-5.6μm) **[A,B,C]** |
| 3. Penetrating Radar and Altimeter (> 20MHz) **[B,C]** | 3. Atmospheric Structure Instrument and Meteorological Package (including a nephelometer/particle counter, an anemometer, and temperature and pressure sensors) **[A]** |
| 4. Mid- to Far-Infrared Spectrometer (5-1000 μm) **[A,B,C]** | 4. Electric Environment and Surface Science Package (including a penetrator and a drill) **[A,B,C]** |
| 5. CosmOrtbitrap – high resolution mass spectrometer (up to 10000amu) **[A,C]** | 5. Radar sounder (> 150MHz) **[B,C]** |
| 6. Icy grain and organic dust analyzer **[A,B,C]** | 6. CosmOrbitrap – Gas Chirality Chromatograph Mass Spectrometer (1-600amu) **[A,B,C]** |
| 7. Plasma suite **[A,B,C]** | 7. Radio science using spacecraft telecom system **[A,B,C]** |
| 8. Magnetometer **[A,B,C]** | 8. Magnetometer **[A,B,C]** |
| 9. Radio Science Experiment **[A,B,C]** | 9. Neutron-activated gamma-ray spectrometer **[B,C]** |
| 10. Sub-Millimeter Heterodyne Receiver **[ A,B,C]** | 10. Seismometer **[B,C]** |
| 11. UV Spectrometer **[A,B,C]** | |

### 5.2.2 Critical issues and technological developments

For long-duration surface missions, solar power is inefficient and radioisotope power sources are the only alternative. In the TSSM concept, MMRTGs or ASRGs using $^{238}$Pu were considered and were to be provided by NASA. Within Europe the radioisotope $^{241}$Am is considered a feasible alternative to $^{238}$Pu and can provide a heat source for small-scale Radioisotope Thermoelectric Generators (RTGs) and Radioisotope Heating Units (RHUs) (Sarsfield et al. 2012), albeit with higher mass. $^{241}$Am exists in an isotopically pure state within stored civil plutonium at reprocessing sites – about 1000kg of $^{241}$Am exists in the civil $PuO_2$ stockpile of the UK and France. A study is underway to design a process that will chemically separate $^{241}$Am (Sarsfield et al. 2012). The development of $^{241}$Am-based RTGs is under consideration by ESA and should be available at high TRL before the proposed Voyage 2050 launch windows.

Drones have never been concretely considered by ESA for planetary exploration. However, their technology seems now mature for application to planetary exploration and is of greatest value for the investigation of remote places such as planetary atmospheres and surfaces. Following the impetus given by NASA by the selection of Dragonfly and the Mars Helicopter Scout (Ingenuity), we advocate that ESA conducts technical analyses expressly dedicated to the feasibility study and technological developments of planetary flying drones. A detailed comparison between the different approaches (one heavily instrumented drone, an air fleet of mini-drones, or a lake lander) will be needed to determine the best option for *in situ* exploration of Titan's atmosphere and surface.

Instrumenting a drone or the lake lander's heat shield with a geophysical and meteorological package (a seismometer and possibly other instruments, such as a drill, a penetrometer, an electrical environment package, an anemometer, a pressure sensor, etc) can also be considered. Instrumenting the probes' heat shield was already considered by TSSM. Such options would require further study to evaluate their feasibility and utility. Finally, support from national agencies will be essential in developing the next generation of highly capable instruments, as well as in pursuing experimental and modeling efforts initiated with Cassini-Huygens, in order to be ready for this next rendez-vous with Titan.

#### 5.2.3 Possibilities for descoping the ESA mission class and opportunities for international collaboration

Mission scenarios within different budget envelopes would include, by increasing order of mission class: an **M-class** mission concept with only an *in situ* element (lake lander, drone, a few mini-drones), but missing all the key science questions related to global processes, an **L-class** mission concept including a Titan orbiter and small, focused *in situ* element(s) (lake probe, a few mini-drones), missing key questions at regional scale, and an **L$^+$-class** (with international collaboration to support the overall architecture and cost) mission concept including a Titan orbiter and at least one ambitious *in situ* element (a flying, 'amphibious'/floating drone), allowing us to address all the fundamental questions summarized in this document.

Should the mission concept be descoped, one could also consider the participation of ESA with another space agency to provide (partial) support for an orbiter including key instruments (CosmOrbitrap, sub-mm spectrometer, near-IR and visible spectral-imager, radar experiment, and small plasma and magnetosphere *in situ* analysis suites) or an *in situ* probe for Titan exploration, in synergy with the incoming NASA mission(s) to Titan. In that sense, for instance, ESA could propose an orbiter to go with Dragonfly from the beginning of its mission and have the octocopter as the *in situ* element. In addition, to assist the TOPS concept being considered by NASA for further feasibility study, ESA could participate by providing a second *in situ* element or the lake lander, or participate to their conception.

Table 2: Detailed list of key questions associated to Goals A, B, and C, of the measurement requirements and proposed Titan orbiter and *in situ* probe instrumentation to address them (in blue and framed in blue when concerning *in situ* probe(s)).

| | *Key measurements to address the question* | *Proposed instrument* |
|---|---|---|
| **A : Titan's atmosphere** | | |
| **A. 1 Chemistry and physical processes in the upper atmosphere (from 500km to 1400km)**<br>• Nature of the dust-plasma interaction and impact on the ionosphere?<br>• Electron balance in Titan's dayside? Which role does the ion transport from dayside play to maintain the nightside ionosphere?<br>• What drives the variability in the thermosphere?<br>• Chemical nature of the macromolecules?<br>• What are the relative contributions of ion-neutral vs. radical reaction pathways?<br>• What is the nature, intensity, and time variability of the source(s) of oxygen in the atmosphere? Does it get efficiently incorporated into molecules? Can biological compounds (amino acids, nucleic bases, etc) or other chemical species with some prebiotic potential be synthesized in the atmosphere?<br>• What is the source of the 1000km supersonic zonal wind? How does it vary with season and interact with the ionosphere?<br>• What are the global dynamics of the upper atmosphere? | Ions and neutrals densities from *in situ* at ≈1000km. Electron density and temperature.<br><br>Negative ion composition and densities.<br><br>Electron intensities.<br><br>Magnetic field vector.<br><br>Vertical profiles of $N_2$, $CH_4$, hydrocarbons, and nitriles.<br><br>Temperature profiles, gravity wave activity.<br><br>Aerosol extinction and particle properties.<br><br>Direct wind measurements from molecular line Doppler shift.<br><br>Temperature profiles, mixing ratio profiles. | **CosmOrbitrap:** *in situ* ion and neutral mass spectrometer, combined with a Pressure gauge for total neutral density [NOTE: INMS good for composition but not total densities]<br>**Mutual Impedence Probe /Langmuir probe**: electron density and temperature [NOTE: to compare with total positive ions and see the departure due to dust-plasma interaction]<br>**Negative ion mass spectrometer:** negative ion mass spectra and densities<br>**Fluxgate magnetometer**: three components of the magnetic field<br>**Electron spectrometer**: electron intensities as a function of energy<br>**UV spectrometer**: temperature, aerosol extinction, particle properties, $N_2$, $CH_4$, hydrocarbons, and nitriles mixing ratio profiles<br>**Sub-mm spectrometer:** temperature, $CH_4$, $CH_3CCH$, nitriles (larger than those observed in UV) mixing ratio profiles. Direct wind measurement from line Doppler shifts. |
| **A.2 Dynamics and chemistry in the middle atmosphere (from 100km to 600km)**<br>• Structure, frequency, and seasonal evolution of planetary wave? Their impact on angular momentum transport, on the haze distribution? What is their relation to the zonal wind field and the polar vortex?<br>• What controls the polar vortex latitudinal extent, how it forms and ends and what is its vertical structure across mesosphere and stratosphere?<br>• What is the aerosol spectral refractive index? Does it change spatially/with season?<br>• Composition of the massive polar stratospheric clouds? | Observation of the signatures of the waves on the haze and cloud images.<br><br>Horizontal and vertical mapping of thermal field, photochemical species, and haze spatial distribution and the wind speed with time.<br><br>Spectra of the aerosol optical depth in IR, visible, and UV plus their spatial and vertical variations.<br><br>IR spectra of the stratospheric clouds. | **Visible and near-IR imager:** nadir and limb viewing<br><br>**Thermal infrared (mid- and far-IR) spectrometer**<br><br>**Sub-mm spectrometer:** nadir and limb viewing<br><br>**UV, vis, IR spectrometer** |
| **A.3 The lower atmosphere: clouds, weather, and methane cycle**<br>• Origin of the atmospheric methane?<br>• $CH_4$ humidity spatial and seasonal variations?<br>• Dynamics, frequency, and precipitation rates of convective methane storms?<br>• Wind speed in the lower troposphere? Over the lakes?<br>• Origin of the zero zonal wind speed at 80km? Its seasonal evolution and consequences on angular momentum exchanges between troposphere and stratosphere? What are the consequences on the haze and trace species transport through this region?<br>• What is the cloud composition?<br>• What is the aerosol spectral refractive index? Does it change spatially/with season? | Monitoring of cloud activity, precipitations, and methane humidity.<br><br>Temperature and wind profiles in the deep atmosphere retrieved from radio occultation observations.<br><br>Solar aureole, remote observation of the column opacity, average aerosol/cloud particles size, spectra.<br><br>Aerosol density and size distribution. Cloud particle size distribution. | **Visible camera and a near-infrared spectrometer**<br><br>**Radio antenna**<br><br>**Imager/spectral radiometer**<br><br>**Nephelometer/particle counter** |
| **B : Titan's geology** | | |
| **B.1 Aeolian features and processes**<br>• What is the precise – not extrapolated – geographic distribution of Titan's aeolian landforms?<br>• What is the precise morphometry of the dunes and does it change with locations on Titan?<br>• What is the wind regime responsible for dunes' and other aeolian landforms' morphology and orientation?<br>• Are dunes and other aeolian landforms still active today? | Global topography and mapping of Titan's surface at decametric spatial resolution - extraction of the aeolian units and accurate. characterization of their geographic distribution, of dune width and spacing statistics, and crest orientations at global scale.<br>High spatial and spectral resolution near-infrared spectra of aeolian features and their geological context.<br>Stereo-imaging at very high spatial resolution.<br>Sampling of the dune material. | **Near-infrared camera and spectrometer**<br>**Penetrating radar and altimeter**<br>**Near-infrared camera and spectrometer**<br>**Electric environment package**<br>**Mini-CosmOrbitrap**<br>**Neutron-activated gamma-ray spectrometer** |
| • What are the origin, sources, and sinks of Titan's sand and can we draw the pathways of sediment transport? Why are there no dunes in Xanadu region?<br>• Composition, grain size, degree of cohesion, and durability of the dune material? | | |
| **B.2 Fluvial features and processes**<br>• What is the precise – not extrapolated – geographic distribution of Titan's aeolian landforms?<br>• What is the precise morphometry of the dunes and does it change with locations on Titan?<br>• What is the wind regime responsible to dunes' and other aeolian landforms' morphology and orientation?<br>• Are dunes and other aeolian landforms still active today?<br>• What are the sources and sinks of Titan's sand? Can we draw the pathways of sediment transport? Why are there no dunes in the Xanadu region?<br>• What is the composition, grain size, degree of cohesion and durability of the dune material? | Global topography and mapping of Titan's surface at decametric spatial resolution - extraction of the fluvial features and accurate characterization of their morphologies.<br>High spatial and spectral resolution near-infrared spectra of fluvial features and their geological context.<br>Stereo-imaging at very high spatial resolution.<br>Sampling of the sediment (e.g. in alluvial fans). | **near-infrared camera and spectrometer**<br>**penetrating radar and altimeter**<br>**near-infrared camera and spectrometer**<br>**radar sounder**<br>**Electric environment package**<br>**Mini-CosmOrbitrap**<br>**Neutron-actived gamma-ray spectrometer** |
| **B.3 Seas and lacustrine features and processes**<br>• What are the shapes of the lacustrine features in the polar regions?<br>• What is the true distribution of sub-kilometer lakes and what does this tell us about lake formation?<br>• How much liquid is stored in the depressions, how do they connect with a subsurface liquid hydrocarbon table, and what is the true total inventory of organics in the polar areas?<br>• What are the exact compositions of the lakes and seas, how and why do they differ?<br>• By which geological processes do the lacustrine depressions and raised ramparts or rims form?<br>• What are the lake seasonal/short timescale changes? | Global topography and mapping of Titan's surface at decametric spatial resolution - extraction of the seas and lakes' features and accurate characterization of their morphologies.<br>Bathymetry.<br>High spatial and spectral resolution near-infrared spectra of seas and lakes and their geological context.<br>Stereo-imaging at very high spatial resolution.<br>Sampling of the shorelines and possibly liquids. | **near-infrared camera and spectrometer**<br>**radar imager**<br>**penetrating radar and altimeter**<br>**near-infrared camera and spectrometer**<br>**radar sounder**<br>**Electric environment package**<br>**Mini-CosmOrbitrap**<br>**Neutron-actived gamma-ray spectrometer** |
| **B.4 Impact craters and Mountains**<br>• What are the relative ages of all of Titan's geologic units?<br>• What is Titan's bedrock/crust composition?<br>• What are the erosion and degradation rates of craters and mountains? What do they reveal about Titan's past and present climatology? What is the reason for the difference in the crater population of Xanadu Regio from other regions on Titan, and in particular for the paucity of craters in Titan's polar regions? | Global topography and mapping of Titan's surface at decametric spatial resolution - crater size distribution as complete as possible towards the small sizes.<br>High spatial and spectral resolution near-infrared spectra of craters and their geological context.<br>Stereo-imaging at very high spatial resolution.<br>Sampling of the craters' rim, floor, and ejecta material. | **near-infrared camera and spectrometer**<br>**radar imager**<br>**near-infrared camera and spectrometer**<br>**Electric environment package**<br>**Mini-CosmOrbitrap**<br>**Neutron-actived gamma-ray spectrometer** |
| **B.5 Exchange processes with a subsurface global ocean**<br>• What are the depth, volume, and composition of the subsurface liquid water ocean?<br>• Is Titan currently, or has it been in the past, cryovolcanically active?<br>• Are there chemical interactions between the ocean, the rock core, and the organic-rich crust?<br>• How did Titan's atmosphere form and evolve with time in connection with the interior? | Constrain the hydrosphere structure, the ocean composition, and the thickness of the outer shell and the HP ice mantle.<br>Measurement of the ratio between radiogenic and non-radiogenic isotopes in noble gases (Ar, Ne, Kr, Xe) in Titan's atmosphere.<br>Search for surface thermal anomalies and sensing the shallow subsurface and detect warmth from old lava flows.<br>Comparison between isotopic ratio in N, H, C, and O-bearing species in the atmosphere (gas and aerosols) and in collected surface materials.<br>Surface sampling and analysis of erupting materials. | **Near-infrared camera and spectrometer**<br>**Radio experiment**<br>**Radar imager, altimeter, and sounder**<br>**High-resolution microwave radiometer**<br>**Magnetometer and plasma package**<br>**CosmOrbitrap**<br>**Seismometer, radio transponder, electric sensors and magnetometer**<br>**Radar sounder**<br>**Mini-CosmOrbitrap** |
| **C : Titan's habitability** | | |
| • What is the nature and quantity of material exchange between the subsurface ocean and the surface? In the past, did a form of life develop in the water pond, formed by cryovolcanism or bolide impacts?<br>• How is the organic material falling from the atmosphere physically/chemically processed at the surface? Does some catalytic path exist for the hydrogenization of acetylene or | Chemical analysis of gases, liquids, and solids (possibility after dissolution in a specific chamber) composing the ground or sea.<br><br>Imaging with spectro-imaging and direct sampling of seas' surface. | **A liquid sampling system / drilling system coupled to an analysis instrument**<br>**Mini-CosmOrbitrap**<br>**Chiral column chromatography instrument** |
| other reactions? How prevalent is water-ice on Titan's surface? What is the depth of organic deposit on the ice (if measurable by drilling and/or radar)?<br>• Does a layer of surfactant (or even thicker deposit) cover the surface of some lakes/maria?<br>• What is the nature of dissolved species in hydrocarbon lakes? Does this liquid environment harbor a chemical reactions network?<br>• Are the molecules present in lakes and evaporites deposits optically active? Can a kind of homochirality be exhibited? | Collection and analysis of materials of all nature belonging to Titan's atmosphere precipitations.<br><br>Imaging in polarized light/IR. Measurements of optical rotation or detection of molecule chirality. | **Imaging system with several polarization filters available**<br><br>**Polarimeter resembling a Laurent polarimeter** |

## 6. Conclusion

The list of outstanding questions pertaining to Titan as a system is lengthy – a legacy of the extraordinarily successful Cassini-Huygens mission, as befits a world perhaps second only to Earth in its level of geologic and atmospheric activity. Such questions are not merely specific to Titan but have much broader and deeper implications for our comprehension of the habitable conditions in the Solar System and beyond. For these reasons, **we recommend the acknowledgement of Titan as a priority within ESA's Voyage 2050 program**, which has as one of the themes identified the icy moons of the giant planets (https://www.esa.int/Science_Exploration/Space_Science/Voyage_2050_sets_sail_ESA_chooses_future_science_mission_themes). We advocate combining efforts, in science and technology, with international agencies to launch a dedicated and ambitious L-class mission to Titan. Our mission concept POSEIDON (Titan POlar Scout/orbitEr and *In situ* lake lander and DrONe explorer) is to perform joint orbital and *in situ* exploration of polar regions of Titan, complementing the timing, location, and scope of the NASA Dragonfly mission. Such a mission architecture will certainly stimulate important technological advances through the challenging new components required for this investigation. The POSEIDON mission concept will allow us to identify key areas for technology development and corresponding development of a technology plan. It is highly recommended, if not mandatory, for the international space agencies to combine efforts and collaborate in order to make such an ambitious endeavour a reality. International participation among ESA, NASA, and other potential partners will play a key role in achieving all the science goals of this mission, which will revolutionize our understanding not only of the Titan system, but of Earth and life's origins in the Solar System, Galaxy, and Universe.

**Conflict of Interest**

The authors declare no conflict of interest.